%% file: main.tex
\documentclass[11pt]{article}

\usepackage{acl}

\usepackage{times}
\usepackage{latexsym}

\input{math_commands}

\usepackage[T1]{fontenc}
\usepackage[utf8]{inputenc}

\usepackage{microtype}

\usepackage{graphicx}
\usepackage{lineno}
\usepackage{hyperref}
\usepackage{url}
\usepackage{listings}
\usepackage{booktabs}       % professional-quality tables
\usepackage{amsfonts}       % blackboard math symbols
\usepackage{amsmath}        % align environment and more
\usepackage{xcolor} % for colored text
\usepackage{nicefrac}       % compact symbols for 1/2, etc.
\usepackage{xcolor}         % colors
\usepackage{xspace}
\usepackage{tabularx}
\usepackage{enumitem}

\usepackage{cleveref}
\usepackage{multirow}

\usepackage{geometry}
\usepackage{array}
\usepackage{wrapfig,booktabs}
\usepackage{subcaption, float}

\usepackage[most]{tcolorbox}
\tcbuselibrary{listings}
\tcbuselibrary{breakable}
\tcbuselibrary{skins}
\title{OmniCode: A Benchmark for Evaluating Software Development Agents}

\author{\textbf{Atharv Sonwane\textsuperscript{1, *}},
  \textbf{Eng-Shen Tu\textsuperscript{2, *}},
  \textbf{Wei-Chung Lu\textsuperscript{3, *}},
  \textbf{Claas Beger\textsuperscript{1, *}},
  \textbf{Carter Larsen\textsuperscript{1}},
  \textbf{Debjit Dhar\textsuperscript{4}},
\\
  \textbf{Simon Alford\textsuperscript{1}},
  \textbf{Rachel Chen\textsuperscript{5}},
  \textbf{Ronit Pattanayak\textsuperscript{1}},
  \textbf{Tuan Anh Dang\textsuperscript{1}},
  \textbf{Guohao Chen\textsuperscript{1}},
  \textbf{Gloria Geng\textsuperscript{1}},
\\
  \textbf{Kevin Ellis\textsuperscript{1}},
  \textbf{Saikat Dutta\textsuperscript{1}}
\\
\\
  \textsuperscript{1}Cornell University,
  \textsuperscript{2}Independent contributor,
  \textsuperscript{3}UC Santa Barbara
  \textsuperscript{4}Jadavpur University
  \textsuperscript{5}New York University
\\
\textsuperscript{*}Equal contribution
\\
\\
}

\begin{document}

\newcommand{\mypara}[1]{\vspace{.03in}\noindent \textbf{#1.}}

\input{macros}

\maketitle
\begin{abstract}
LLM-powered coding agents are redefining how real-world software is developed.
To drive the research towards better coding agents, we require challenging benchmarks that can rigorously evaluate the ability of such agents to perform various
software engineering tasks. However, popular coding benchmarks such as HumanEval and SWE-Bench focus on narrowly scoped tasks such as competition programming and patch generation. 
In reality, software engineers have to handle a broader set of tasks for real-world software development. 
To address this gap, we propose \tool, a novel software engineering benchmark that contains a broader and more diverse set of task categories beyond code or patch generation.
Overall, \tool contains \totaltasks tasks spanning three programming languages -- Python, Java, and C++ -- and four key categories: bug fixing, test generation, code review fixing, and style fixing.
In contrast to prior software engineering benchmarks, the tasks in \tool are (1) manually validated to eliminate ill-defined problems, and (2) synthetically crafted or recently curated to avoid data leakage issues, presenting a new framework for synthetically generating diverse software tasks from limited real-world data.
We evaluate \tool with popular agent frameworks such as SWE-Agent and show that while they may perform well on bug fixing for Python, they fall short on tasks such as Test Generation and in languages such as C++ and Java.
For instance, SWE-Agent achieves a maximum of \CppDeepSeekTest~with \deepseek on C++ Test Generation.
\tool aims to serve as a robust benchmark and spur the development of agents that can perform well across different aspects of software development.
Code and data are available at \url{https://github.com/seal-research/OmniCode}.

\end{abstract}

\section{Introduction}

The future impact of AI-automated software development will be far-ranging:
beyond building and improving apps, AI will help us write more comprehensive test suites,
perform and respond to code review suggestions, 
enforce nuanced programming guidelines, and automate many other tasks that are part of the software development life cycle.
Research on AI software development demands good benchmarks, both to measure progress and to expand the scope of problem statements.
However, AI coding benchmarks today, such as SWE-Bench~\citep{jimenez2024swebench}, CodeContests~\citep{li2022competition}, and HumanEval~\citep{chen2021evaluatinglargelanguagemodels}, are too narrow in scope to spur progress on automating the full spectrum of software development tasks, instead focusing on isolated tasks such as competition programming and patch generation.

\begin{figure}[!htb]
\centering
\includegraphics[scale=0.34]{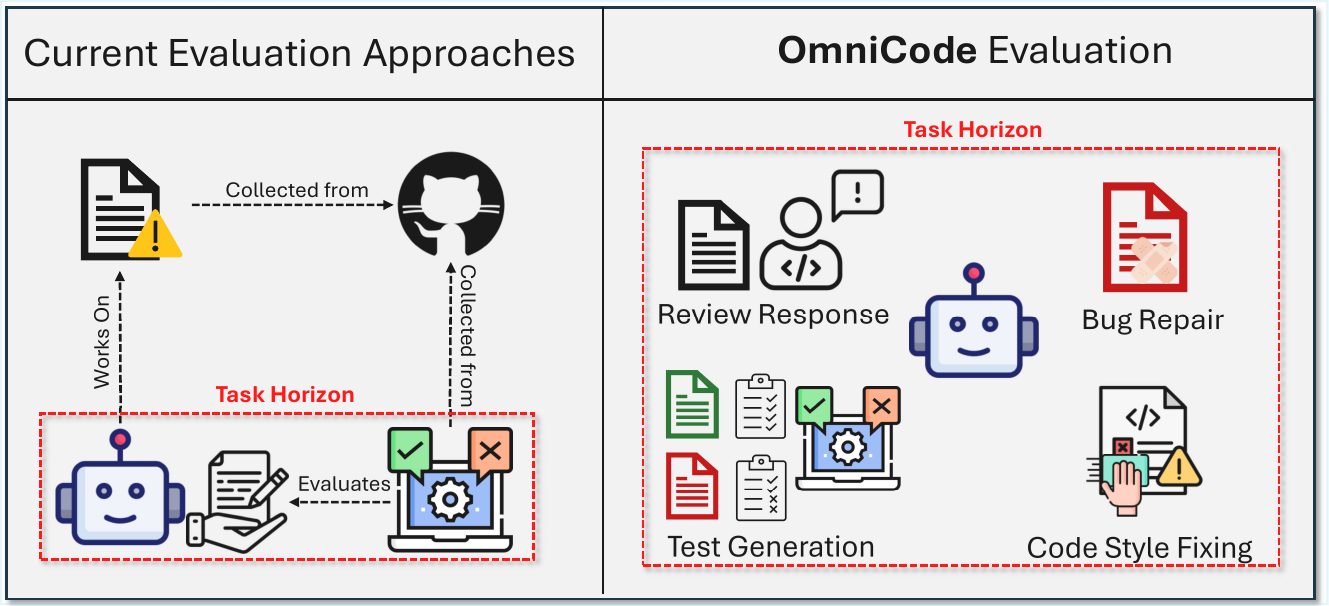}
\caption{\tool synthetically builds multiple tasks out of a base dataset to holistically evaluate software engineering agents. Four different types of tasks that we consider: Bug fixing, test generation, responding to code review, and enforcing style guidelines.}
\label{fig:overview}
\end{figure}

\mypara{\tool} To address this gap, we introduce a new benchmark for evaluating generative AI coding assistants, which we call \tool. Our new benchmark is based on the insight that software development involves a heterogeneous range of tasks and problem-solving activities for which generative AI can be brought to bear (see Figure~\ref{fig:overview}).
We consider four such software development tasks:
\begin{enumerate}[leftmargin=*]
    \item \textbf{Addressing issues, such as bug fixes and feature requests.}
    This is a staple of software engineering benchmarks~\citep{jimenez2024swebench,silva2024repairbench,rashid2025swe}, because it evaluates the ability of an LLM coding agent to autonomously resolve real-world repository-level issues.
    
    \item \textbf{Writing software tests.}
    We evaluate the ability of LLM agents to write their own tests, measuring progress toward closing the loop of both generating and checking patches. 
    We create a dataset of \textit{bad patches} which implement incorrect fixes. In addition to passing on the correct fix, we ensure that the generated tests must fail on the bad patches (emulating mutation testing~\cite{jia2010analysis}), making the evaluation of tests more robust.

    \item \textbf{Responding to code review.}
    Coding agents today act in partnership with human engineers, providing initial drafts of a patch, which a human engineer then critiques.
    We compile a dataset of partly-correct patches paired with code-review feedback on how to best correct them.

    \item \textbf{Enforcing style guidelines.} Code style is important for conforming to project-specific norms and ensuring safe coding practices.
    We task the agent with fixing selections of style violations. These include functions with high complexity, unsafe direct access of union members or assigning outputs of functions that return null.
    
\end{enumerate}

We build our benchmark by bootstrapping off existing benchmarks such as SWE-Bench~\citep{jimenez2024swebench} and Multi-SWE-Bench~\citep{zan2025multiswebenchmultilingualbenchmarkissue}, along with collecting additional issues from popular open-source repositories and across three popular languages: Python, Java, and C++.
We build on top of this real-world data with LLM and tool-based augmentation to create different task types. 
For supporting test generation, we generate \textit{bad patches} with LLMs, and for the review-fix tasks we similarly generate possible code reviews using LLMs.
For style review, we use language-specific style-checking tools to create tasks.
In total, our dataset comprises 494 issues from \totalrepos repositories and \totaltasks benchmark tasks in total, with \pythontasks Python tasks, \javatasks Java tasks, and \cpptasks C++ tasks.
% Repos: python: 14, java: 8, cpp: 5

\mypara{Results} We evaluate the widely used SWE-Agent~\citep{yang2024swe} with SOTA models spanning a range of providers and sizes (\gemini, \claude, \gptmini, \deepseek, and \qwen) on our dataset.
We also evaluate another agent, Aider~\citep{aiderai} (with \gemini), which is a pipeline-based agent unlike SWE-Agent.
We find that our benchmark challenges even the most modern systems, but it is not intractable.
% \Fix{add numbers for all tasks and model names}
% Specifically, \sweagent achieves a maximum of \JavaDeepSeekTest~on test generation across all three languages. On Review-Response it achieves a maximum of \PyDeepSeekReview~on Python. For Style-Fixing, while agents perform well on Python, they do not perform as well on Java and C++.
Specifically, \sweagent achieves its strongest bug-fixing results with \claude on Python (\PyClaudeBug) but drops sharply on Java (\JavaDeepSeekBug\ with \deepseek) and C++ (\CppDeepSeekBug\ with \deepseek).  
Test generation is the hardest category: the best score across all three languages is only \CppDeepSeekTest\ (\deepseek on C++). On Review-Response, \claude reaches \PyClaudeReview\ on Python, while the best 
Java and C++ scores are \JavaDeepSeekReview\ and \CppDeepSeekReview\ (both \deepseek). For Style-Fixing, agents perform well on Python (up to \PyClaudeStyle\ with \claude) but lag on Java (\JavaClaudeStyle) and C++ (\CppClaudeStyle).
We also observe that SWE-Agent generally outperforms Aider, with Aider performing significantly worse on C++, scoring 3–5× lower across all four task types.
% (e.g., \CppAiderBug\ vs \CppGeminiBug\ on bug-fixing, \CppAiderStyle\ vs \CppGeminiStyle\ on style).

\mypara{Contributions}
We make the following contributions in this work:
\begin{enumerate}[leftmargin=*,noitemsep,topsep=0in]
    \item We develop \tool, a benchmark assessing for diverse types of software engineering activities, comprising \totaltasks tasks total.
    \item We present strategies for synthetically creating diverse interactive tasks to evaluate agents from collected static real-world data.
    \item We perform empirical evaluation of \sota LLM-agent systems on the benchmark, revealing specific areas where LLM agents fall especially short, particularly in test generation and style fixing. 
\end{enumerate}

\section{Related Work}

\mypara{LLM coding benchmarks}
Early evaluation of the LLM code synthesis focused on standalone programming problems with unit-test-based functional correctness~\citep{chen2021evaluatinglargelanguagemodels, Austin2021ProgramSW, Hendrycks2021MeasuringCC}.
Work such as CodeMind~\citep{Liu2024CodeMindAF} benchmarks semantic reasoning about code behavior, while CoCoNUT~\citep{11028290} targets structural execution of small code blocks across programming paradigms.
SWE-Bench~\citep{jimenez2024swebench} introduced the paradigm of benchmarking the ability of LLM agents to resolve real-world GitHub issues, yielding much follow-up work~\citep{miserendino2025swelancerfrontierllmsearn, jain2024r2e, aleithan2024swebenchenhancedcodingbenchmark, rashid2025swe,zan2024swe}, adding support for more repositories, languages and improved data quality by including more rigorous checks.
Multi-SWE-Bench~\citep{zan2025multiswebenchmultilingualbenchmarkissue} extended the SWE-Bench collection paradigm beyond Python to multiple languages, but restricted to bug-fixing.
These bug-fixing benchmarks are related to work from the software engineering community benchmarking program repair tools such as Defects4J~\citep{Just2014Defects4JAD} and BugsInPy~\citep{Widyasari2020BugsInPyAD}.
Other benchmarks evaluate security-focused tasks, such as CyberGym~\cite{wang2026cybergym}, CVEBench~\cite{cvebench} and CWE-Bench~\cite{li2024iris}.
While these benchmarks focus on bug-fixing or security, \tool introduces pipelines for creating new task types to more holistically evaluate software development.

Recently works such as SWE-Smith~\citep{yang2025swesmithscalingdatasoftware} and BugPilot~\citep{sonwane2025bugpilotcomplexbuggeneration} have shown promise in synthetically generating bugs to create training data for coding agents.
\tool makes use of synthetic data in the context of evaluation to go beyond bug-fixing.
We introduce pipelines to augment real-world data to support new task types such as test generation and responding to reviews.

\mypara{Test generation}
Recently, Mündler et al. proposed SWT-Bench~\citep{mundler2024swt} that transforms the instances in SWE-Bench to test generation tasks. 
Each task involves generating tests such that they fail on the buggy version of code and pass with the fixed version, e.g., the gold patch. 
TestEval~\citep{wang2024testeval} is another recent benchmark for evaluating test generation capabilities of LLMs, focusing on evaluating single programs instead of entire repositories.
The test generation task in \tool builds on these works while strengthening the oracle: its test-generation task evaluates candidate tests not only on the gold fix but also against multiple plausible bad patches, reducing the chance that vacuous tests pass trivially and better measuring discriminative power.

\mypara{LLM Agents for code} \cite{yang2024swe} introduces SWE-Agent, one of the first agent-based systems for SWE tasks. Many other such agents have been proposed since, such as AutoCodeRover~\citep{zhang2024autocoderover}, Agentless~\citep{xia2024agentless}, OpenHands~\citep{Wang2024OpenHandsAO} and Aider~\citep{aiderai}. 
Agents such as QLCoder~\cite{Wang2025QLCoderAQ} interact with powerful static analysis tools to find security vulnerabilities.
These developments have rapidly improved the quality of agents, indicated by their impressive scores on the SWE-Bench benchmark. We hope that \tool encourages a more well-rounded evaluation of such agents.

\section{Benchmark Construction}

\tool consists of different task types, the creation of which involves applying various task-specific augmentations to real-world software data.
For the bug-fixing, test generation, and review response task types, each instance in our benchmark is based on a pull request that was raised to resolve an issue in a GitHub repository.
The pull request and its associated metadata (such as the issue it resolved, the patch it introduced) constitute what we call a \emph{base instance}.
Using this base instance, we can generate the data required to support different task types, such as generating bad patches to support test generation or code reviews to support review fixing.
For the style fix task type, we create instances by running style tools on open source repositories.

\subsection{Collection of real-world data from GitHub} \label{sec:collection}

\begin{table}
  \vspace{-0.5\baselineskip} 
  \centering
  \footnotesize
  \caption{Combined patch statistics by language}
  \label{tab:combined_lang_stats}
  \begin{tabular}{l | r r r}
    \toprule
    \textbf{Metric} & \textbf{Python} & \textbf{C++} & \textbf{Java} \\
    \midrule
    \multicolumn{4}{l}{\textit{Patch statistics}} \\
    \cmidrule(lr){1-4}
    Patches        & \PatchCountPy        & \PatchCountCpp        & \PatchCountJava        \\
    Complexity     & \PatchComplexityPy   & \PatchComplexityCpp   & \PatchComplexityJava   \\
    Lines added    & \PatchAddPy          & \PatchAddCpp          & \PatchAddJava          \\
    Lines removed  & \PatchRemovePy       & \PatchRemoveCpp       & \PatchRemoveJava       \\
    \cmidrule(lr){1-4}
    \addlinespace[0.3em]

    \multicolumn{4}{l}{\textit{Test statistics}} \\
    \cmidrule(lr){1-4}
    Patches        & \TestCountPy         & \TestCountCpp         & \TestCountJava         \\
    Complexity     & \TestComplexityPy    & \TestComplexityCpp    & \TestComplexityJava    \\
    Lines added    & \TestAddPy           & \TestAddCpp           & \TestAddJava           \\
    Lines removed  & \TestRemovePy        & \TestRemoveCpp        & \TestRemoveJava        \\
    \midrule
    \addlinespace[0.3em]

    \multicolumn{4}{l}{\textit{Bad Patch and Review statistics}} \\
    \cmidrule(lr){1-4}
    Patches        & \BadPatchCountPy      & \BadPatchCountCpp      & \BadPatchCountJava      \\
    Complexity     & \BadPatchComplexityPy & \BadPatchComplexityCpp & \BadPatchComplexityJava \\
    Lines added    & \BadPatchAddPy        & \BadPatchAddCpp        & \BadPatchAddJava        \\
    Lines removed  & \BadPatchRemovePy     & \BadPatchRemoveCpp     & \BadPatchRemoveJava     \\
    Review size    & \ReviewSizePy         & \ReviewSizeCpp         & \ReviewSizeJava         \\
    \bottomrule
  \end{tabular}
  \vspace{-0.75\baselineskip}
\end{table}

We curate a multi-language dataset by collecting data from open source GitHub repositories, along with selecting instances from existing benchmarks: SWE-Bench-Verified and Multi-SWE-Bench.
When curating pull requests, we follow a similar selection strategy to
\cite{jimenez2024swebench}, including PRs that (1) resolve an issue and (2) introduce at least one test.
We further filter instances manually to ensure data quality (see Appendix~\ref{app:data} for details).

% We also curate instances sourced from SWE-Bench-Verified and Multi-SWE-Bench, 
% filtering out non-functional instances where dependencies could not be installed properly or where pre-existing tests would not pass.
% We supplement this with a small number of additional repositories (5 repos for Python, 2 repos for Java, and \Fix{TODO} for C++).

% By extending coverage to Java and C++ in addition to Python, our dataset broadens evaluation beyond the Python-centric scope of SWE-Bench, providing a more realistic and comprehensive benchmark for assessing software engineering agents across ecosystems. While we choose Python, Java, and C++ languages in this work, we can easily expand to other languages in the future, following the same approach.

To enable agents to interact with an instance by executing code, we build containerized environments for each instance.
The environment is made up of the state of the repository at the time of the pull request, as well as dependencies that need to be installed so that code can be executed properly.
We manually determine the dependencies required by inspecting requirements and documentation.
To verify that the correct dependencies have been identified, we execute the test suite of the repository to check if the tests can be run without errors.

Our combined dataset of base instances comprises \pythonbftasks Python, \cppbftasks C++, and \javabftasks Java instances (\totalbftasks in total), spanning \totalrepos diverse repositories across machine learning and scientific libraries (e.g., Scikit-learn, Sympy), systems libraries (e.g., Fmt, Simdjson), and large-scale frameworks (e.g., Django, Logstash, Jackson, Mockito). 
Our data pipelines are easily extensible to other languages for future work.

\subsection{Task Details}
\label{sec:TaskEv}
In Table~\ref{tab:combined_lang_stats}, we report quantitative statistics over these gold patches to characterize task difficulty across languages.
Beyond raw patch size, we introduce a composite complexity metric to better capture structural edit difficulty:
\begin{equation}
\label{eq:patch_complexity}
\Delta \text{Files}
+ \Delta \text{Hunks}
+ \Delta \text{Lines} / 10.
\end{equation}

This metric incorporates the files modified, number of code hunks and lines changed in a diff.
It jointly reflects the breadth of codebase interaction, edit fragmentation, and textual change magnitude.
Using this measure, we observe a consistent difficulty ordering across languages:
\textbf{C++ > Java > Python}.
This ordering aligns with downstream agent performance trends reported in Section~\ref{sec:eval}, suggesting that patch complexity is a meaningful proxy for task difficulty.
In the following, we describe the details of how each of our four task types is set up, along with the evaluation procedures. 

\subsubsection{Task: Resolving Issues}
\label{sec:bugfixing}
Resolving GitHub issues has become a standard approach for evaluating the capabilities of large language models (LLMs) in the software engineering domain. 
We use our \textit{base instances} directly for this task, following \cite{jimenez2024swebench}, we provide the issue description and a set of tests that distinguish between the pre- and post-fix repository states. 
The agent is tasked with generating a patch based on the issue, which is evaluated against tests that transitioned from failing to passing due to the ground truth fix, as well as against previously passing tests to ensure no regressions are introduced.
Our benchmark goes beyond prior work by unifying instances from SWE-Bench-Verified, Multi-SWE-Bench, as well as including manually verified additional instances (details in Appendix~\ref{app:data}).

\subsubsection{Task: Test Generation}
\label{sec:testgen}

\begin{figure}
\centering
\includegraphics[scale=0.5]{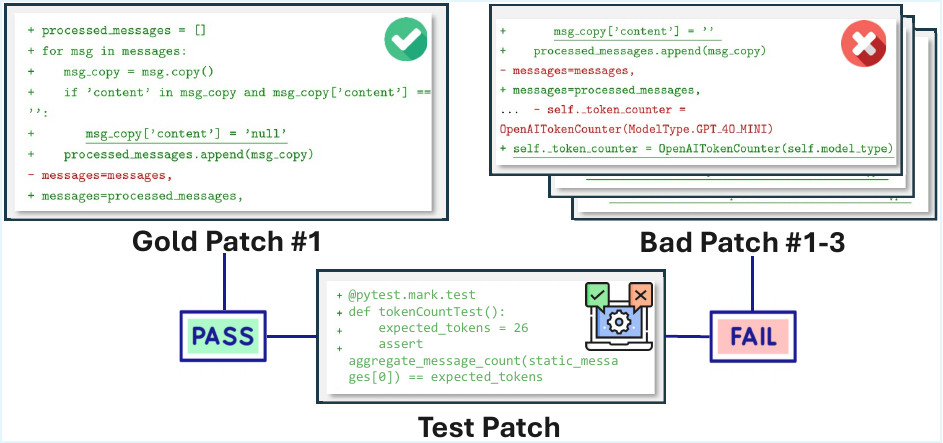}
\caption{In the Test Generation task, we evaluate proposed test patch against both the ground truth (gold) patch, as well as several meaningful, but incorrect, bad patches. A test is only considered correct if it passes for the gold patch, but fails for all bad patches.
}
\label{fig:test-gen}
\end{figure}

Writing meaningful tests is a key aspect of software engineering, and by focusing on this underexplored skill, we aim to evaluate the model's ability to reason about code behavior and generate effective test cases.
To assess the quality of a candidate test for a given issue, we use the ground truth fix for that issue (termed as the \emph{gold patch}) along with a set of what we define as \emph{bad patches}. A bad patch is a plausible but incorrect attempt at resolving the issue—one that contains no obvious syntax errors and remains relevant to the problem description. This setup presents a more realistic and challenging evaluation scenario compared to existing approaches, which typically rely only on the pre- and post-PR repository states. 

To generate bad patches that are plausible but incorrect, we collect failed attempts from an agent (Agentless~\citep{xia2024agentless}) tasked with fixing issues as usual.
A well-generated test should be able to distinguish the correct solution from failure modes exhibited by a collection of such failed attempts.
Since relying on a single model for this exercise often results in some instances without bad patches of this sort, we use a mix of models (Gemma 2 9B, Qwen2.5 Coder 32B Instruct, Llama 3 8B Instruct, and GPT-4.1-nano). 
% producing most of the bad patches for Java and C++, and GPT-4.1-nano producing most of the bad patches for Python.\Fix{what about other languages?}
To ensure diversity of bad patches for better evaluation of generated tests, we aim to generate 3 patches for each instance.
Some instances have fewer bad patches due to invalid generations or cases where an incorrect fix could not be sampled after repeated attempts.
Instances with no bad patches are excluded from the test generation task.

To ensure that generated tests can be evaluated thoroughly, it is important to have bad patches that cover a diverse set of failure modes.
We experiment with another approach for generating bad patches, prompting an LLM to perturb the correct patch to introduce commonly found bugs.
We sample multiple completions, filtering to keep patches that are actually incorrect.
The relevant prompt can be found in Appendix~\ref{app:prompts}.
We only use this approach for Python instances, producing one bad patch for each instance using Gemini-2.5-Flash.

For this task, the agent is prompted with the issue text and asked to generate one or more test cases to be added to the test suite. 
The generated tests are then evaluated: if they pass on the ground truth patch but fail on all bad patches, they are considered successful. 
If the set of generated tests do not meet both criteria, they are considered a failure.
We also reuse the bad patches in an additional task related to code review (\S~\ref{sec:codereview}).

\subsubsection{Task: Responding to Code Review}
\label{sec:codereview}

\begin{figure}[t]
  \centering
  \includegraphics[width=0.8\columnwidth]{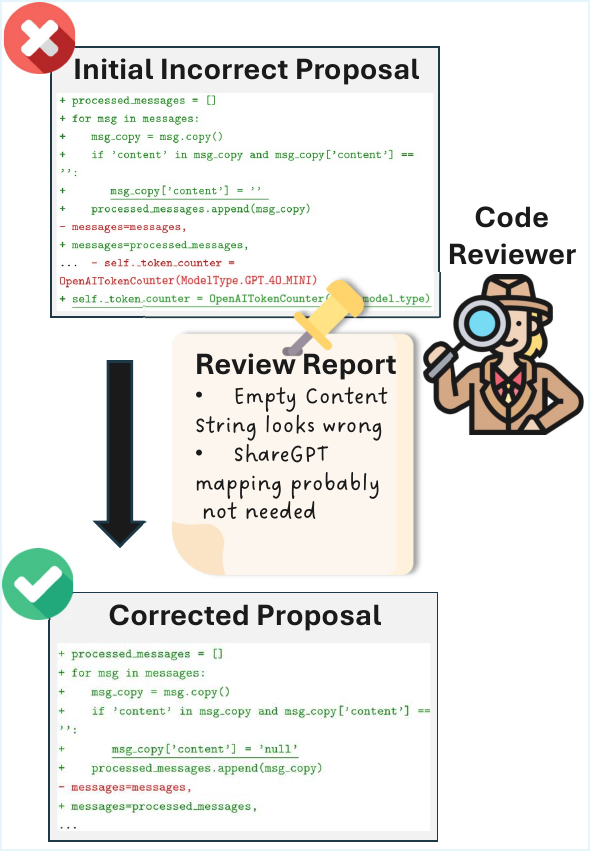}
  \caption{In the task of responding to Code Review, an initial incorrect patch is provided, which contains a meaningful attempt at solving a given problem. This attempt is then reviewed by a human or an LLM, and a review report is generated. Utilizing this report, the LLM is tasked with correcting the initial approach, which is then validated with the normal testing suite.}
  \label{fig:review}
\end{figure}

Developers often iterate over multiple proposed solutions in a pull request until they fulfill all the necessary requirements.
Often, such incorrect proposals are met with a corresponding \emph{review}, explaining inadequacies in approach or implementation.
For the code review response task in our benchmark, we create reviews by prompting an LLM (\gemini) with both the bad patch (from the previous section), along with the correct patch and problem description, and asking it to come up with instructions for how the bad patch should be fixed.
We find that the simple prompt presented in Appendix~\ref{app:prompts} produces reviews that are informative but do not give away the complete solution on manual inspection of a subset of generated reviews.
During evaluation, we present the agent with the problem description, incorrect fix (bad patch), along with the review, tasking it with refining the solution.
% While the adaptation of existing functionality to enable this use case is minor, we believe this is a promising avenue for research into fully autonomous work on code issues, interacting with external feedback, and starting from potentially corrupted states.

\subsubsection{Task: Fixing Code Style Issues}
Language models are trained on a wide range of code -- varying not only in functionality but also in quality.
So, tasks that assess the ability of models to adhere to style guidelines represent a natural extension of evaluation.
In this task, the model is not expected to fix a functional bug but to resolve a given set of style issues. 
This task is particularly appealing because it can be adapted to user-specific needs by incorporating custom guidelines or organization-specific rules.

We construct a dataset of style errors for all 28 repositories (14 Python, 3 C++, and 11 Java) in \tool. 
We start by using the language-specific tools (\texttt{pylint} for Python, \texttt{clang-tidy} for C++, and \texttt{PMD} for Java) to generate a list of all style violations in the repository, including errors, warnings, and convention violations.
We then aggressively prune trivial rules (e.g., missing newlines or extra whitespaces). 
The full list of included style violations is present in Appendix~\ref{sec:style_rules_app} in Table~\ref{tab:py-warnings} (Python), Table~\ref{tab:java-warnings} (Java), and Table~\ref{tab:cpp-warnings} (C++).
We then group errors by files they occur in and construct \pythonsftasks Python, \cppsftasks C++, and \javasftasks Java instances, with each instance containing on average $\sim$ 9 style errors.

During evaluation for each instance, the style errors as given by the linter tool are passed to the agent, which is tasked with resolving them.
After applying the patch generated by the agent, we re-run the style tool over the whole repository to determine how many of the previous errors were resolved and how many new ones were introduced.
To ensure that while correcting style issues, the agent does not inadvertently break existing functionality, we also run the repository's test suite.
Finally, we compute a \textbf{Style-Fixing Score} for the patch as given below —

\begin{equation}
\label{eq:style_score_definition}
\begin{cases}
\max\!\left(
\dfrac{\text{resolved} - \text{new}}{\text{original}},\, 0
\right),
& \text{if all tests pass} \\[8pt]
0,
& \text{otherwise}
\end{cases}
\end{equation}

Patches that result in test failures are assigned a score of 0. 
If the existing functionality is preserved, then we compute a metric that measures what proportion of the original issues were resolved, penalizing newly created issues.
Instances where more issues are introduced than resolved result in a score of 0.
This design ensures the metric is bounded between 0 and 1, taking into account correctness as well as newly created issues.

% \begin{figure}[t]
% \begin{subfigure}[t]{0.6\linewidth}
%   %\vspace{-0.5\baselineskip}
%   \centering
%   \includegraphics[scale=0.3]{CodeReview2.pdf}
%   \caption{In the task of responding to Code Review, an initial incorrect patch is provided, which contains a meaningful attempt at solving a given problem. This attempt is then reviewed by a human or an LLM, and a review report is generated. Utilizing this report, the LLM is tasked with correcting the initial approach, which is then validated with the normal testing suite.}
%   \label{fig:review}
%   %\vspace{-0.75\baselineskip}
% \end{subfigure}

%\begin{subfigure}[t]{0.4\linewidth}
\begin{figure}[t]
\centering
\includegraphics[width=0.85\columnwidth]{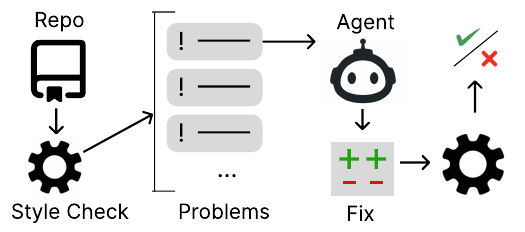}
\caption{
In the Style Fix task, we first create task instances by running a style check tool on the whole repository and grouping local issues into instances. The agent is then tasked with fixing these instances, with the proposed fix being evaluated by running the style checker again.
}
\label{fig:style-fix}
\end{figure}
%\end{subfigure}
%\end{figure}

\section{Evaluation of LLM Coding Agents on \tool}
\subsection{Experimental Setup}
To evaluate our benchmark, we select two state-of-the-art agent frameworks: \sweagent{}~\citep{yang2024swe} and Aider~\citep{aiderai}.
To enable agents to interact with the instances, we provide them with containerized environments as described in \S~\ref{sec:collection}.
We pass in the issue description as the initial task statement for Bug-Fixing. For Test-Generation, Review-Response, and Style-Fixing, we prepare task-specific prompts that provide context and instructions. These are detailed in Appendix~\ref{app:prompts}.
We use the default settings for \sweagent and adjust the per instance cost limit to \$2.0. 
We perform Aider evaluations with \gemini, setting the timeout to 20 minutes and retry-attempts to 3.

\mypara{LLMs used} For our main evaluation with \sweagent, we choose five \sota LLMs from different model families: \gemini, \claude, \deepseek, \gptmini, and \qwen. For \aider, we only evaluate using \gemini to limit costs.

\label{sec:eval}

\begin{table*}[h]
\centering
\footnotesize
\caption{SWE-Agent Performance on OmniCode across languages and models}
\label{tab:sweagent_langs}
\begin{tabular}{l l c c c c}
\toprule
\textbf{Language} & \textbf{Model} & \textbf{Bug-Fixing} & \textbf{Test-Generation} & \textbf{Review-Response} & \textbf{Style-Fixing} \\ 
\midrule
\multirow{5}{*}{Python}   
  & Gemini-2.5-Flash    & \PyGeminiBug      & \PyGeminiTest      & \PyGeminiReview      & \PyGeminiStyle \\ 
  & Claude-4.6-Sonnet   & \textbf{\PyClaudeBug}      & \PyClaudeTest      & \textbf{\PyClaudeReview}      & \textbf{\PyClaudeStyle} \\
  & GPT-5-mini          & \PyGPTMiniBug     & \PyGPTMiniTest     & \PyGPTMiniReview     & \PyGPTMiniStyle \\
  & DeepSeek-V3.1       & \PyDeepSeekBug    & \textbf{\PyDeepSeekTest}    & \PyDeepSeekReview    & \PyDeepSeekStyle \\  
  & Qwen3-32B           & \PyQwenBug        & \PyQwenTest        & \PyQwenReview        & \PyQwenStyle \\ 
\midrule
\multirow{5}{*}{C++}    
  & Gemini-2.5-Flash    & \CppGeminiBug     & \CppGeminiTest     & \CppGeminiReview     & \CppGeminiStyle \\
  & Claude-4.6-Sonnet   & \CppClaudeBug     & \CppClaudeTest     & \CppClaudeReview     & \textbf{\CppClaudeStyle} \\
  & GPT-5-mini          & \CppGPTMiniBug    & \CppGPTMiniTest    & \CppGPTMiniReview    & \CppGPTMiniStyle \\
  & DeepSeek-V3.1       & \textbf{\CppDeepSeekBug}   & \textbf{\CppDeepSeekTest}   & \textbf{\CppDeepSeekReview}   & \CppDeepSeekStyle \\
  & Qwen3-32B           & \CppQwenBug       & \CppQwenTest       & \CppQwenReview       & \CppQwenStyle \\
\midrule
\multirow{5}{*}{Java}   
  & Gemini-2.5-Flash    & \JavaGeminiBug    & \JavaGeminiTest    & \JavaGeminiReview    & \JavaGeminiStyle \\
  & Claude-4.6-Sonnet   & \JavaClaudeBug    & \JavaClaudeTest    & \JavaClaudeReview    & \textbf{\JavaClaudeStyle} \\
  & GPT-5-mini          & \JavaGPTMiniBug   & \JavaGPTMiniTest   & \JavaGPTMiniReview   & \JavaGPTMiniStyle \\
  & DeepSeek-V3.1       & \textbf{\JavaDeepSeekBug}  & \textbf{\JavaDeepSeekTest}  & \textbf{\JavaDeepSeekReview}  & \JavaDeepSeekStyle \\
  & Qwen3-32B           & \JavaQwenBug      & \JavaQwenTest      & \JavaQwenReview      & \JavaQwenStyle \\
\bottomrule
\end{tabular}
\end{table*}

% \begin{table*}[h]
% \centering
% \footnotesize
% \caption{SWE-Agent vs Aider Comparison}
% \label{tab:gemini_agents}
% \begin{tabular}{l l c c c c}
% \toprule
% \textbf{Language} & \textbf{Agent} & \textbf{Bug-Fixing} & \textbf{Test-Generation} & \textbf{Review-Response} & \textbf{Style-Fixing} \\
% \midrule
% \multirow{2}{*}{Python} 
%     & SWE-Agent (Gemini-2.5-Flash)  & \textbf{\PyGeminiBug}  & \textbf{\PyGeminiTest}  & \textbf{\PyGeminiReview}  & \textbf{\PyGeminiStyle} \\
%     & Aider (Gemini-2.5-Flash)      & \PyAiderBug            & \PyAiderTest            & \PyAiderReview            & \PyAiderStyle \\

% \cmidrule(lr){1-6}
% \multirow{2}{*}{C++}    
%     & SWE-Agent (Gemini-2.5-Flash)  & \textbf{\CppGeminiBug} & \textbf{\CppGeminiTest} & \textbf{\CppGeminiReview} & \textbf{\CppGeminiStyle} \\
%     & Aider (Gemini-2.5-Flash)      & \CppAiderBug           & \CppAiderTest           & \CppAiderReview           & \CppAiderStyle \\
% \cmidrule(lr){1-6}
% \multirow{2}{*}{Java}   
%     & SWE-Agent (Gemini-2.5-Flash)  & \JavaGeminiBug         & \textbf{\JavaGeminiTest} & \textbf{\JavaGeminiReview} & \textbf{\JavaGeminiStyle} \\
%     & Aider (Gemini-2.5-Flash)      & \textbf{\JavaAiderBug} & \JavaAiderTest           & \JavaAiderReview           & \JavaAiderStyle \\
% \bottomrule
% \end{tabular}
% \end{table*}

\begin{table}[t]
\centering
\footnotesize
\caption{SWE-Agent vs Aider (using \gemini)}
\label{tab:agents}
\setlength{\tabcolsep}{0.04in}
\begin{tabular}{l l c c c c}
\toprule
\textbf{Language} & \textbf{Agent} & \textbf{BF} & \textbf{TG} & \textbf{RR} & \textbf{SF} \\
\midrule
\multirow{2}{*}{Python} 
    & SWE-Agent  & \textbf{\PyGeminiBug}  & \textbf{\PyGeminiTest}  & \textbf{\PyGeminiReview}  & \textbf{\PyGeminiStyle} \\
    & Aider      & \PyAiderBug            & \PyAiderTest            & \PyAiderReview            & \PyAiderStyle \\

\cmidrule(lr){1-6}
\multirow{2}{*}{C++}    
    & SWE-Agent  & \textbf{\CppGeminiBug} & \textbf{\CppGeminiTest} & \textbf{\CppGeminiReview} & \textbf{\CppGeminiStyle} \\
    & Aider      & \CppAiderBug           & \CppAiderTest           & \CppAiderReview           & \CppAiderStyle \\

\cmidrule(lr){1-6}
\multirow{2}{*}{Java}   
    & SWE-Agent  & \JavaGeminiBug         & \textbf{\JavaGeminiTest} & \textbf{\JavaGeminiReview} & \textbf{\JavaGeminiStyle} \\
    & Aider      & \textbf{\JavaAiderBug} & \JavaAiderTest           & \JavaAiderReview           & \JavaAiderStyle \\
\bottomrule
\end{tabular}
\end{table}

\subsection{Observed Performance across Tasks}

We present the results of evaluating SWE-Agent with a range of \sota LLMs in Table~\ref{tab:sweagent_langs}.
We find that while a \sota system like SWE-Agent excels on some tasks like Style Fixing in Python, there are many shortcomings in other tasks.
We observe that it struggles with Test-Generation, an essential skill for assisting humans and self-verification. All LLMs struggle here across languages, with the maximum performance across all three languages being \CppDeepSeekTest\ (\deepseek on C++).
Both tools also suffer disproportionately in C++, which correlates with our analysis in Section~\ref{sec:TaskEv} regarding the complexity of C++ bugs over other bugs in our benchmark.

\begin{table}[h]
\centering
\footnotesize
\caption{Correlation of Bug-Fixing with other tasks}
\label{tab:bf_corr}
\setlength{\tabcolsep}{0.05in}
\begin{tabular}{lrrr}
\toprule
\textbf{Language} & \textbf{Review vs Bug} & \textbf{Test vs Bug} & \textbf{Style vs Bug} \\
\midrule
Python & 0.925 & 0.702 & 0.800 \\
C++    & 0.966 & 0.733 & 0.461 \\
Java   & 0.871 & 0.858 & 0.276 \\
\midrule
\textbf{Average} & \textbf{0.921} & \textbf{0.764} & \textbf{0.512} \\
\bottomrule
\end{tabular}
\end{table}

\mypara{Correlation across tasks types} As reported in Table~\ref{tab:bf_corr}, we find that performance of different models on bug-fixing is strongly correlated to review-response (pearson coeff = 0.921) and weakly correlated to test generation (pearson coeff = 0.764). 
This does not hold for style-review however (pearson coeff = 0.512), where \gemini performs as good as or better than \deepseek on Style-Fix, despite DeepSeek consistently outperforming Gemini on Bug-Fix.
We find these observations to be generally true for Aider, too, albeit slightly weaker with further details in Appendix~\ref{sec:task_correlation}.

% \mypara{Cost-constrained inference} As shown in \ref{tab:sweagent_langs}, Claude-4.6-Sonnet performs strongly on Python tasks, showcasing the dominance of frontier models, but such performance does not carry over to Java and C++. A key reason is the fixed \$2 per-instance cost constraint used across all experiments. Under this budget, Claude exhibits substantially lower patch generation rates on higher complexity tasks, see \ref{tab:sweagent_patch_generation_rate} in the Appendix, which directly suppresses its resolve rate because instances that do not produce a patch before the budget is exhausted are counted as unresolved. This effect is particularly apparent in Java and C++, where average task complexity is higher compared to Python, see \ref{tab:combined_lang_stats}. This suggests that Claude's lower performance on Java and C++ should not be interpreted as weaker problem-solving ability, but as an interaction between model cost, task complexity, and imposed budget constraint.

\mypara{Cost-constrained inference} Claude-4.6-Sonnet leads on Python but trails on Java and C++ (Table~\ref{tab:sweagent_langs}), which we attribute to the fixed \$2 per-instance
budget. On the higher-complexity Java and C++ tasks (Table~\ref{tab:combined_lang_stats}), Claude frequently exhausts its budget before emitting a patch (Table~\ref{tab:sweagent_patch_generation_rate}), and such runs are counted as unresolved. Its lower scores reflect the cost--complexity--budget interaction rather than
weaker problem-solving ability.
Compute the resolve rate for only the instances where patches were generated, discarding instances where the budget was exceeded: 
$\textit{Score} = \textit{Resolved} / \textit{PatchesGenerated}$ in Table~\ref{tab:sweagent_langs_capability}, we see \claude leading across tasks.

\begin{table}[h]
\centering
\footnotesize
\caption{Style Review Analysis (models from Table~\ref{tab:sweagent_langs})}
\label{tab:style_scores}
\setlength{\tabcolsep}{0.05in}
\begin{tabular}{llrrr}
\toprule
\textbf{Language} & \textbf{Model} & \textbf{Fix Rate} & \textbf{Error Ratio} & \textbf{Style-Fix}\\
\midrule
\multirow{5}{*}{Python}
& Gemini             & \textbf{\PyGeminiFix}   & \PyGeminiErr             & \PyGeminiStyle \\
& Claude  & \PyClaudeFix            & \PyClaudeErr             & \textbf{\PyClaudeStyle} \\
& GPT                & \PyGPTMiniFix           & \PyGPTMiniErr            & \PyGPTMiniStyle \\
& DeepSeek           & \PyDeepSeekFix          & \textbf{\PyDeepSeekErr}  & \PyDeepSeekStyle \\
& Qwen               & \PyQwenFix              & \PyQwenErr               & \PyQwenStyle \\
\midrule
\multirow{5}{*}{C++}
& Gemini             & \textbf{\CppGeminiFix}  & \CppGeminiErr            & \CppGeminiStyle \\
& Claude  & \CppClaudeFix           & \textbf{\CppClaudeErr}   & \textbf{\CppClaudeStyle} \\
& GPT                & \CppGPTMiniFix          & \CppGPTMiniErr           & \CppGPTMiniStyle \\
& DeepSeek           & \CppDeepSeekFix         & \CppDeepSeekErr          & \CppDeepSeekStyle \\
& Qwen               & \CppQwenFix             & \CppQwenErr              & \CppQwenStyle \\
\midrule
\multirow{5}{*}{Java}
& Gemini             & \textbf{\JavaGeminiFix} & \textbf{\JavaGeminiErr}  & \JavaGeminiStyle \\
& Claude  & \JavaClaudeFix          & \JavaClaudeErr           & \textbf{\JavaClaudeStyle} \\
& GPT                & \JavaGPTMiniFix         & \JavaGPTMiniErr          & \JavaGPTMiniStyle \\
& DeepSeek           & \JavaDeepSeekFix        & \JavaDeepSeekErr         & \JavaDeepSeekStyle \\
& Qwen               & \JavaQwenFix            & \JavaQwenErr             & \JavaQwenStyle \\
\bottomrule
\end{tabular}
\end{table}

\mypara{New errors introduced during Style-Fixing} Apart from using Equation~\ref{eq:style_score_definition} as a general metric that takes code functionality into account, we also check how the agents perform on Style-Fixing tasks with
$Fix Rate = resolved / original$
and
$Error Ratio = (unresolved + new) / original$.
Based on the results shown in Table~\ref{tab:style_scores}, we can see that while the Fix Rate is relatively high across models and languages, this metric alone overstates the style-fixing reliability of coding agents. The Error Ratio exposes a tendency for style-fixing attempts to introduce new style errors to the code, particularly for structurally complex languages like Java and C++. Python shows the lowest error ratios, which implies that most fixes resolve style violations without substantial side effects. Java and C++ have significantly higher error ratios, reflecting the fragility of style-only transformations in larger codebases. Qualitative analysis of the Style-Fixing tasks and error types is presented in Appendix~\ref{style_review_analysis}.

\mypara{Providing Reviews helps solve issues}
We hypothesized that providing structured feedback would improve performance on tasks by guiding the agent.
We sampled instances with gold patches of higher complexity to generate reviews via the procedure in Section~\ref{sec:codereview} for the Review-Response task.
Comparing performance on this subset, we observe that Review-Response consistently resolved more unique instances than Bug-Fixing (without reviews).
For Java, it uniquely resolved 15 instances versus Bug-Fixing's 4, a pattern that held for C++ (4 vs. 2) and Python (22 vs. 20). 
The seemingly contradictory raw scores in Table~\ref{tab:sweagent_langs} for Python (\PyDeepSeekBug~Bug-Fixing vs. \PyDeepSeekReview~Review-Response) are explained by the non-review instances being comparatively easier, with a high 65.1\% resolution rate.

\subsection{Performance Comparison of Agentic and Pipeline-Based Approaches}
We compare a widely used agentic approach (\textit{SWE-Agent}) with a pipeline-based approach (\textit{Aider}) to assess the strengths and weaknesses of both paradigms. As shown in Table~\ref{tab:agents}, \textit{SWE-Agent} consistently outperforms Aider across most programming languages and task types when evaluated on OmniCode using Gemini-2.5-Flash. For Python, \textit{SWE-Agent} achieves higher performance in bug-fixing, test-generation, and review-response, reflecting its stronger reasoning and synthesis capabilities. 
In C++, \emph{Aider} performs substantially worse, while \textit{SWE-Agent} maintains modest but consistent gains, particularly in test-generation and review-response. One potential reason is that C++ tasks in OmniCode require more interactive reasoning and iterative error analysis, involving multiple compile-run cycles and complex dependency handling. \emph{Aider's} pipeline-oriented design may thus struggle with such trial-and-error-intensive workflows.
Overall, these findings indicate that while \textit{\aider} remains competitive on less interactive or simpler tasks, \sweagent demonstrates greater robustness and adaptability to complex, multi-stage software engineering problems, particularly those requiring sustained reasoning and integration of feedback. These results highlight \tool's ability to differentiate between interaction-intensive and procedural tasks, providing a nuanced view of how agentic and pipeline systems handle varying levels of task complexity and reasoning demand.

\begin{table*}[h]
  \centering % Centers the table
  \footnotesize % Use smaller font size as in the example
  \caption{Bug-Fixing - Avg. Complexity Score.}
  % The \label command MUST come AFTER the \caption.
  \label{table:bugfixing_complexity}
  \setlength{\tabcolsep}{0.05in}
  % New table structure with 8 columns, using \multirow
  \begin{tabular}{l l rrrrrr}
    \toprule % Top line from booktabs
    % Bold headers, with stacking for long ones
    \textbf{Model} & \textbf{Language} & \multicolumn{1}{c}{\textbf{\begin{tabular}[c]{@{}c@{}}Gold Avg\\Complexity\end{tabular}}} & \textbf{\begin{tabular}[c]{@{}c@{}}Model Avg\\Complexity\end{tabular}} & \multicolumn{1}{c}{\textbf{\begin{tabular}[c]{@{}c@{}}Resolved\\Gold Avg\\Complexity\end{tabular}}} & \multicolumn{1}{c}{\textbf{\begin{tabular}[c]{@{}c@{}}Resolved\\Model Avg\\Complexity\end{tabular}}} & \multicolumn{1}{c}{\textbf{\begin{tabular}[c]{@{}c@{}}Unresolved\\Gold Avg\\Complexity\end{tabular}}} & \multicolumn{1}{c}{\textbf{\begin{tabular}[c]{@{}c@{}}Unresolved\\Model Avg\\Complexity\end{tabular}}} \\
    \midrule % Middle line from booktabs
    \multirow{3}{*}{Gemini-2.5-Flash} & Python & 7.07 & 299.28 & 5.35 & 5.28 & 8.13 & 484.67 \\
    & Java & 19.24 & 9.75 & 6.69 & 12.32 & 19.67 & 19.24 \\
    & C++ & 47.55 & 195.1 & 8.07 & 6.28 & 38.26 & 252.31 \\
    \midrule
    \multirow{3}{*}{Claude-4.6-Sonnet} & Python & 7.07 & 259.35 & 5.29 & 4.08 & 11.00 & 941.29 \\
    & Java & 19.24 & 5.75 & 5.94 & 4.28 & 21.70 & 7.68 \\
    & C++ & 47.55 & 13.14 & 11.58 & 17.18 & 50.70 & 12.53 \\
    \midrule
    \multirow{3}{*}{GPT-5-mini} & Python & 7.07 & 165.56 & 4.30 & 4.05 & 9.55 & 390.18 \\
    & Java & 19.24 & 983.12 & 6.51 & 4.24 & 21.95 & 1186.70 \\
    & C++ & 47.55 & 603.39 & 18.48 & 90.91 & 43.92 & 767.78 \\
    \midrule
    \multirow{3}{*}{Deepseek-v3.1} & Python & 7.07 & 12.08 & 5.22 & 5.35 & 9.46 & 21.60 \\
    & Java & 19.24 & 12.91 & 6.47 & 7.07 & 26.49 & 15.84 \\
    & C++ & 47.55 & 104.63 & 9.21 & 32.13 & 54.29 & 123.18 \\
    \midrule
    \multirow{3}{*}{Qwen3-32b} & Python & 7.07 & 464.93 & 5.77 & 4.32 & 7.49 & 642.09 \\
    & Java & 19.24 & 4.76 & 5.26 & 2.7 & 24.28 & 5.08 \\
    & C++ & 47.55 & 140.96 & 5.00 & 4.75 & 46.37 & 148.22 \\
    \bottomrule % Bottom line from booktabs
  \end{tabular}
\end{table*}

\subsection{Patch Complexity as a Latent Factor in Agent Performance}
\begin{figure}[!htb]
\centering
    \includegraphics[width={0.4\textwidth}]{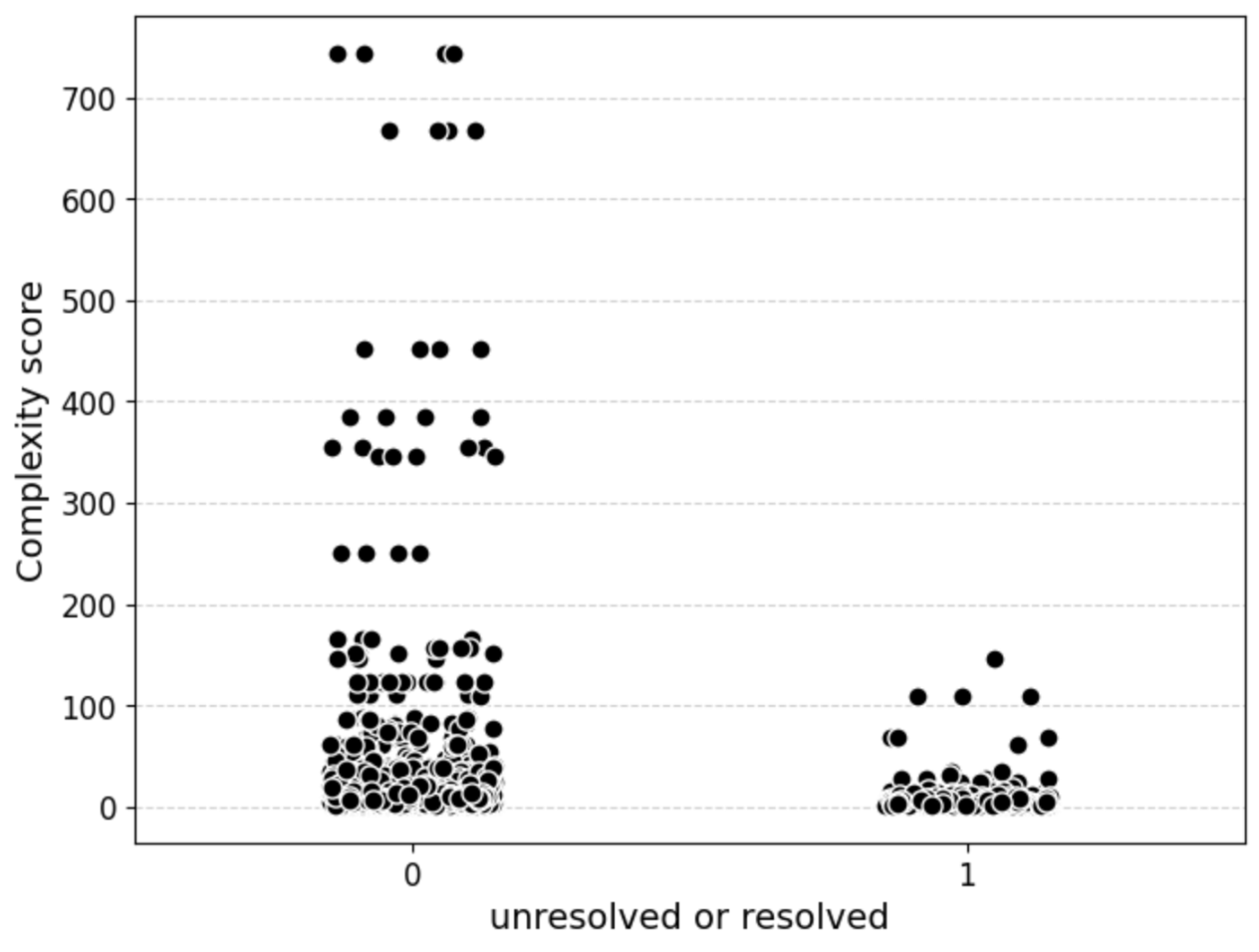}
    \caption{Distribution of patch complexity scores for resolved versus unresolved instances.}
    \label{fig:complexity-distribution}
\end{figure}

Using the complexity metric (Equation~\ref{eq:patch_complexity}), we analyze agent-generated outputs against task difficulty and observe a clear separation between resolved and unresolved instances. According to Figure~\ref{fig:complexity-distribution}, the resolved cases are tightly clustered at low complexity, indicating strong performance on small, localized fixes, while unresolved instances exhibit a heavy-tailed distribution with substantially higher and more volatile complexity. Aggregate statistics for the bug-fixing task as an example (Table~\ref{table:bugfixing_complexity}) quantitatively confirm this trend and illustrate that failures are characterized not only by larger edits but also by unstable patch construction that deviates significantly from Gold solutions. 
We also report detailed complexity statistics indicating that higher patch complexity consistently correlates with lower resolution rates. 
Overall, these results highlight patch complexity as a key latent difficulty factor in software agent benchmarks—benchmarks dominated by low-complexity fixes may overstate agent capability, while higher-complexity instances more reliably expose limitations in long-horizon reasoning and codebase navigation.

% This metric accounts for the breadth of changes (files and hunks) and the depth of modifications (line count). As illustrated in $\ref{fig:complexity-distribution.png}$ (see Appendix), there is a distinct divergence in complexity between resolved and unresolved instances. Specifically, unresolved patches exhibit significantly higher complexity scores, suggesting a negative correlation between structural redundancy and resolution success. Detailed data in \ref{table:bugfixing_complexity} and \ref{table:reviewfix_complexity} reveal that while the ground truth (Gold) patches complexity shows C++ (47.55) $>$ Java (19.24) $>$ Python (7.07), model behaviors vary significantly. DeepSeek-V3.1 demonstrated the highest performance, maintaining average complexity scores closest to the Gold standard. In contrast, other models displayed huge volatility when failing to resolve a bug. For instance, GPT-5-mini’s unresolved Python patches reached an average complexity of 390.18, a nearly 55-fold increase over the Gold average (7.07). Sometimes, successful resolutions are often more concise than human solutions; for example, DeepSeek’s resolved Python patches averaged a complexity of 5.35 (compared to Gold's 7.07). Additionally, while Python failures often result in numerous edits, unresolved Java patches remain minimal (e.g., Qwen3-32B averaged 5.08 vs. Gold 19.24); this pattern is language-dependent. 

\subsection{Impact of including Bad Patches}

\begin{figure}
    \centering
    \vspace{-0.5em}
    \includegraphics[width=0.45\textwidth]{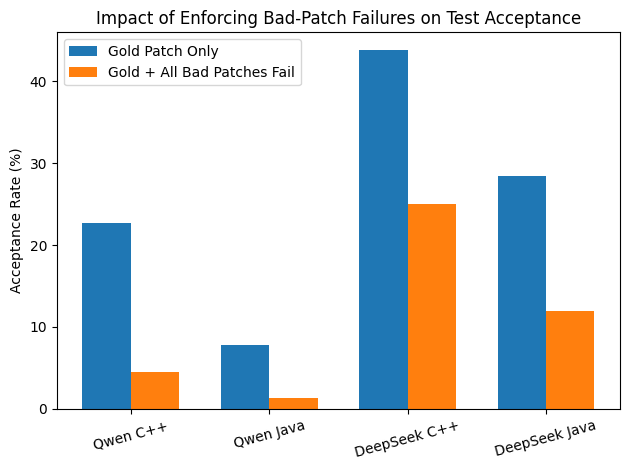}
    \vspace{-0.5em}
    \caption{Comparing agent performance at test generation if evaluating with only the gold patch versus using failing bad patches as a criteria too. Including bad patches leads to more robust results.}
    \label{fig:badpatchinclusion}
\end{figure}

Tests generated by LLM agents need to be evaluated for robustness, making sure that they can discriminate various edge cases.
Prior work~\cite{mundler2024swt} evaluates generated tests solely by relying on fixed and un-fixed versions for a given pull request.
\tool aims to be more robust in its evaluation by ensuring that the generated tests also fail multiple \textit{bad patches}, which characterize different ways in which the bug may be fixed incorrectly.

To quantify the effects of including bad patches, we compare the agent performance if evaluating with only the gold patch as the only criteria versus including failing of the bad patches as a success criteria.
From Figure~\ref{fig:badpatchinclusion}, we can see that metrics based solely on gold-patch success dramatically overestimate a model's testing capability.
In the analysis of success for Qwen and DeepSeek results, test cases would have been accepted at a higher rate if bad-patch failures were not required (e.g., Qwen C++ would be 22.7\% instead of 4.5\%, Qwen Java would be 7.79\% instead of 1.3\%, DeepSeek C++ would be 43.8\% instead of 25\%, and DeepSeek Java would be 28.4\% instead of 20.9\%). 
This gap highlights that many generated tests capture superficial behaviors rather than the underlying program semantics. 
By enforcing that gold patches pass and all bad patches fail, we obtain a far more realistic assessment of test quality, one that reflects a model's ability to differentiate correct logic from subtly incorrect implementations, a critical requirement for trustworthy automated testing.

\section{Conclusion}

We present a novel benchmark for multi-faceted evaluation of coding agents made up of four different tasks across three languages.
Our evaluation of two agent paradigms with an extensive set of models demonstrates various areas for improvement of current coding agents, including consistent test generation, robust handling of style problems as well as improving proficiency at languages beyond Python.
Detailed analysis shows how providing reviews can improve issue resolution and the differences between agentic (SWE-Agent) and pipeline (Aider) based approaches.
We hope that our evaluation and analysis along with the benchmark itself will provide new avenues of improving agents at different aspects of software development.

\section{Limitations and Future Work}
Our work broadens the evaluation of LLMs across diverse software engineering tasks, but it is still far from capturing the full scope of real programming. In practice, developers work across multiple languages, interact with configuration systems, profile and optimize code, and engage in natural language discussions for design and planning. While our benchmark is a step toward more comprehensive assessment, expanding heterogeneous task coverage remains essential.

We are extending \tool along two axes: (1) additional languages beyond Python, Java, and C++, and (2) new task categories such as security vulnerability remediation and code migration. Cross-language translation (e.g., C to Rust) is both challenging and high-impact, while security tasks demand deeper system-level reasoning than current models reliably exhibit. Increasing diversity along these dimensions will enable more realistic and robust evaluation of LLMs and agentic systems.

\bibliography{bibliography}
\appendix

\section{Analysis of Bad Patches and Reviews}
\begin{figure}[H]
\centering
\includegraphics[scale=0.3]{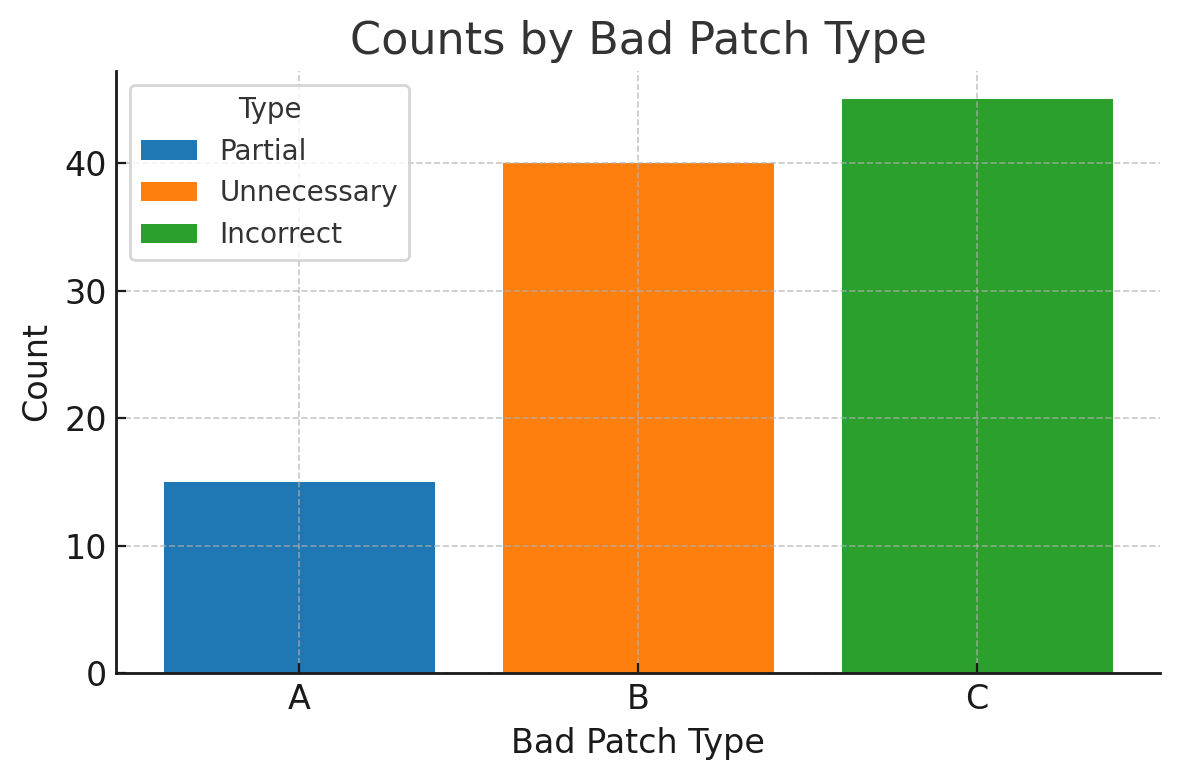}
\caption{Categorization of bad patches.}
\label{fig:bp-cat}
\end{figure}
\begin{figure}[H]
\centering
\includegraphics[scale=0.4]{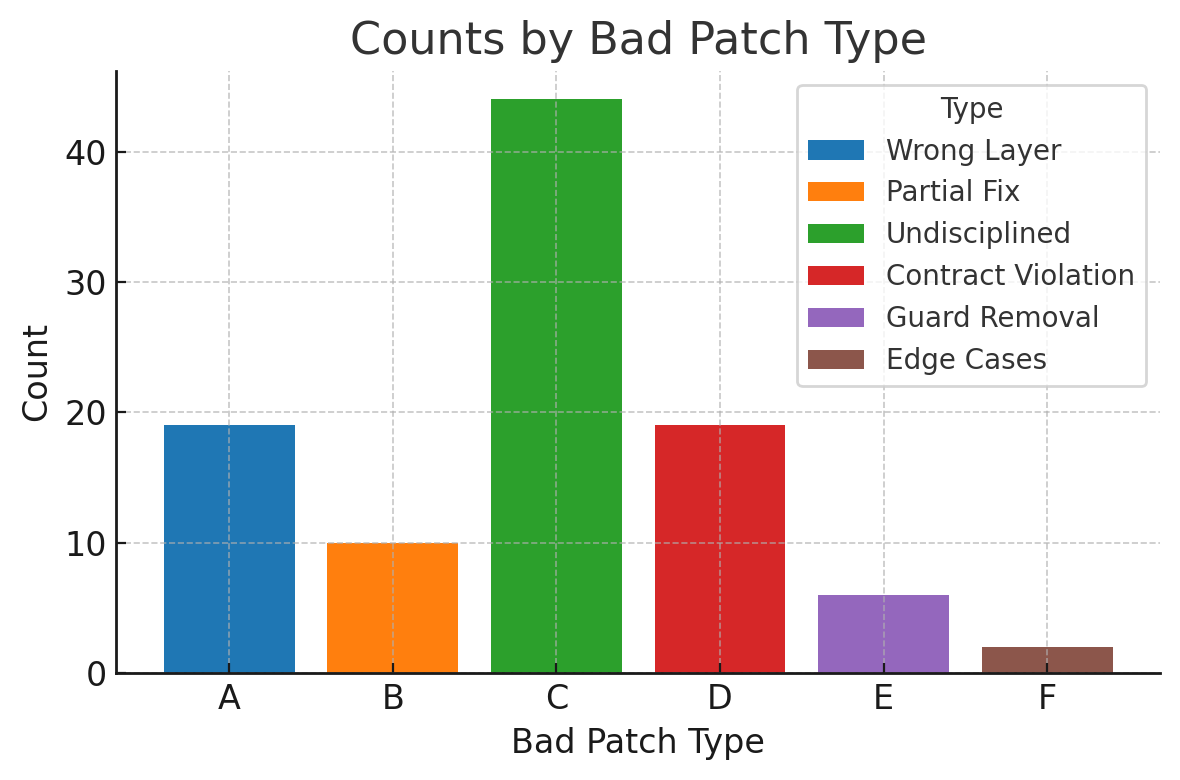}
\caption{Categorization of bad patches.}
\label{fig:bp-cat1}
\end{figure}
\begin{figure}[H]
\centering
\includegraphics[scale=0.4]{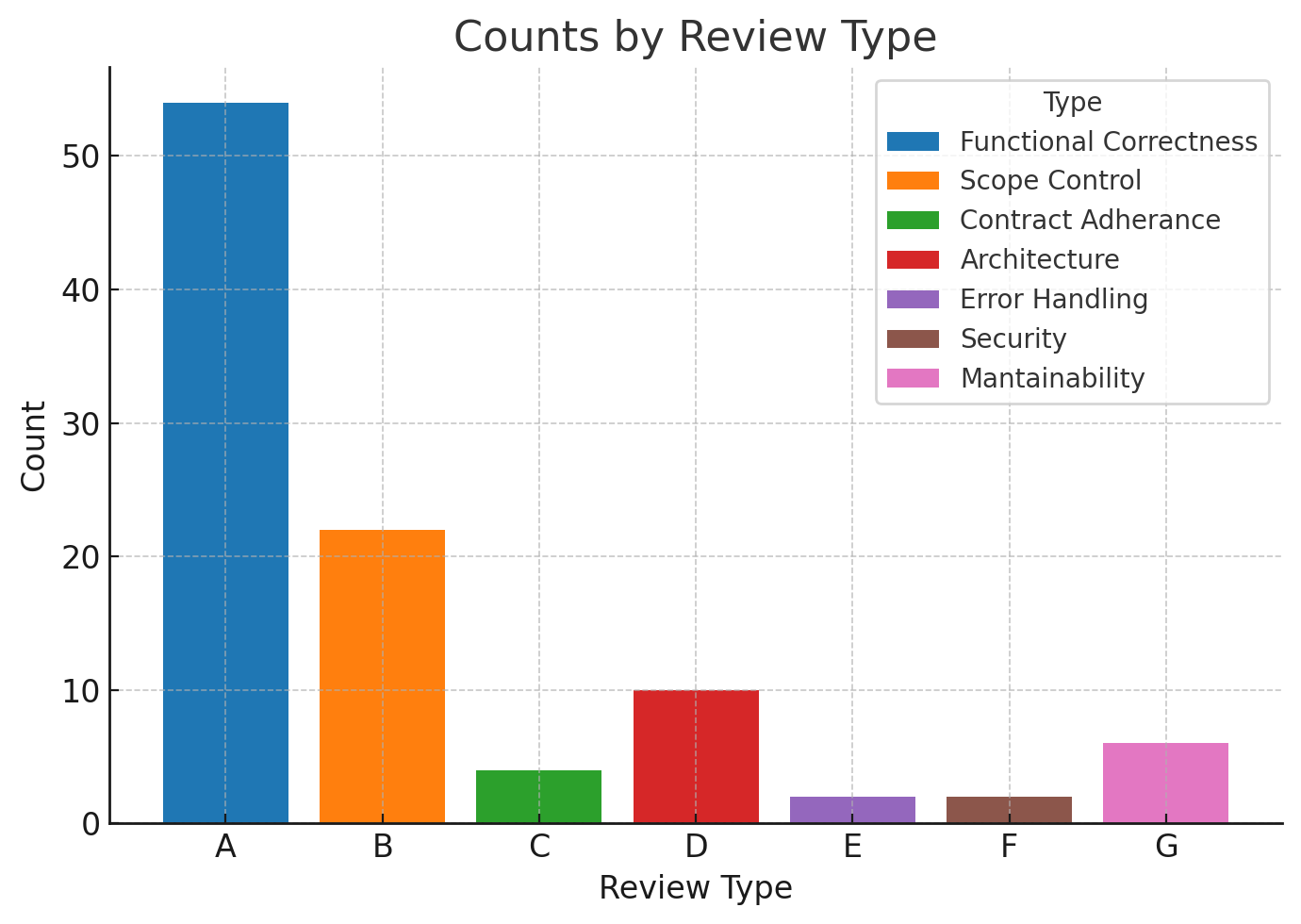}
\caption{Categorization of reviews.}
\label{fig:rev-cat}
\end{figure}

To understand the distribution of bad patches generated by our pipeline, we categorize a sample of 100 Python bad patches with results displayed in Figure~\ref{fig:bp-cat}. The categorization is performed by prompting an LLM with descriptions of the category along with the problem description, bad patch, and correct patch for the instance. 

We observe that bad patches are distributed across a range of types, with most of them being “Undisciplined”, that is, patches that make more spurious changes than necessary. There are also a significant number of bad patches in the “Wrong Layer “ and “Contract Violation” categories.

Another way to understand bad patches is to categorize them according to whether they are “Partial” (the attempted fix is partially correct), "Unnecessary" (the patch makes spurious changes), or “Incorrect” (the fix approach is incorrect). We observe that the majority of bad patches are due to incorrect approach at making the fix. These patches are useful to include in the dataset as they characterize probable failure modes that existing tests may not account for.

To understand the distribution of reviews generated by our pipeline, we categorize a sample of 100 Python reviews with results displayed in Figure~\ref{fig:rev-cat}. The categorization is performed by prompting an LLM with descriptions of the category along with the problem description, bad patch, correct patch, and review for the instance.

Descriptions used to categorize bad patches - 

\begin{enumerate}[label=\textbf{\Alph*.}]
  \item \textbf{Wrong-layer fix / misdiagnosed root cause} \\
  \textit{Description:} The change targets the wrong component or symptom instead of the source of truth. Signals include modifying outputs instead of inputs, tweaking helpers when call sites or flags need changes, relying on attributes/settings that are never wired, or making comment-only/no-op changes.

  \item \textbf{Partial fix / incomplete coverage} \\
  \textit{Description:} Only a subset of affected paths, formats, or call sites is fixed; others remain broken. Typical signs include updating JSON but not XML, adjusting PRAGMA but not SELECT, fixing one code path while an equivalent exists elsewhere, or forgetting to update generated/runtime artifacts.

  \item \textbf{Process hygiene and change discipline failures} \\
  \textit{Description:} The patch mixes unrelated edits (scope creep), alters tests to match a broken implementation, includes merge artifacts or duplicate code, or introduces syntax/typo/runtime errors (duplicate args, unreachable code). These complicate review and often obscure regressions.

  \item \textbf{Contract/invariant violations or Abstraction/API misuse} \\
  \textit{Description:} Changes break explicit or implicit invariants or requirements. Examples include violating “single-column subquery,” making non-atomic multi-step writes, changing multiplication order in non-abelian contexts, keeping multi-column projections inside IN subqueries, bypassing APIs or type contracts, or hardcoding internals. Also includes changing established behavior (defaults, tuple shapes, ordering, observable semantics) without justification or migration.

  \item \textbf{Guard/safety-net removal or inversion} \\
  \textit{Description:} Removing or flipping checks, caches, or validation that protect correctness/security. Indicators include deleting \texttt{is\_active} or \texttt{has\_usable\_password} checks, removing parent\_link validation, dropping inverse/caching assignments, or disabling/inverting critical conditionals.

  \item \textbf{Edge cases, normalization, and type/representation assumptions} \\
  \textit{Description:} Logic fails on uncommon values or conflates representations. Examples: treating \texttt{None} as the only “empty” (ignoring \texttt{''}), mishandling NaN/Inf or undefined semantics, missing lowercase exponent parsing, not rechecking length after mutation, confusing PATH vs PATH\_INFO/script prefixes, or choosing wrappers/proxies that break expected type behavior. Includes overfitted regexes/parsers, missing named groups, unhandled array-indexed dispatch, naive SQL interpolation, missing escaping, off-by-one slices, or wrong encodings/BOM handling.
\end{enumerate}

We perform similar analysis for generated reviews and observe that the vast majority of reviews are to do with improving functional correctness. There are also reviews that discuss “Scope Control” and “Architecture”.
Descriptions used to categorize reviews -

\begin{enumerate}[label=\textbf{\Alph*.}]
  \item \textbf{Functional correctness (logic, control flow, edge cases)} \\
  \textit{Description:} Ensure the fix implements the intended behavior with correct conditions, boundaries, ordering/precedence, and return values. Catch logic/sign errors, unreachable code, inverted conditions, and other correctness issues.

  \item \textbf{Scope control and change isolation} \\
  \textit{Description:} Keep the patch tightly focused on the reported issue. Revert incidental edits, avoid broad refactors, and limit changes to the affected component or backend.

  \item \textbf{API and data contract adherence} \\
  \textit{Description:} Preserve public/internal interfaces, data shapes, and semantics. Avoid breaking consumers, changing return types, or altering documented behavior without coordination.

  \item \textbf{Design/architecture alignment and plumbing} \\
  \textit{Description:} Apply changes in the correct layer (e.g., model vs.\ view), respect separation of concerns, and route control flags/state through the call chain so policies are enforced where needed. Prefer non-breaking or backward-compatible design alternatives.

  \item \textbf{Error and exception handling} \\
  \textit{Description:} Catch and handle expected failures at the correct layer; convert errors to appropriate no-ops or fallbacks. Avoid swallowing unexpected exceptions or leaking internal errors.

  \item \textbf{Security and standards/protocol compliance} \\
  \textit{Description:} Use correct security checks (authz/authn, permission models), avoid unsafe operations (escaping, URL handling), and comply with specified protocols. 
\end{enumerate}

\section{Patch Generation Rate and Capability Within Budget}

\begin{table*}[h]
  \centering
  \footnotesize
    \caption{SWE-Agent Patch Generate Rate across models and tasks}
    \label{tab:sweagent_patch_generation_rate}
    \begin{tabular}{l l c c c c}
    \toprule
    \textbf{Language} & \textbf{Model} & \textbf{Bug-Fixing} & \textbf{Test-Generation} & \textbf{Review-Response} & \textbf{Style-Fixing} \\ 
    \midrule
    \multirow{5}{*}{Python} 
      & Gemini-2.5-Flash  & \PyGeminiPatchBug    & \PyGeminiPatchTest    & \PyGeminiPatchReview    & \PyGeminiPatchStyle \\ 
      & Claude-4.6-Sonnet & \PyClaudePatchBug    & \PyClaudePatchTest    & \PyClaudePatchReview    & \PyClaudePatchStyle \\
      & GPT-5-mini        & \PyGPTMiniPatchBug   & \PyGPTMiniPatchTest   & \PyGPTMiniPatchReview   & \PyGPTMiniPatchStyle \\
      & DeepSeek-V3.1     & \PyDeepSeekPatchBug  & \PyDeepSeekPatchTest  & \PyDeepSeekPatchReview  & \PyDeepSeekPatchStyle \\
      & Qwen3-32B         & \PyQwenPatchBug      & \PyQwenPatchTest      & \PyQwenPatchReview      & \PyQwenPatchStyle \\
    \midrule
    \multirow{5}{*}{C++}    
      & Gemini-2.5-Flash  & \CppGeminiPatchBug   & \CppGeminiPatchTest   & \CppGeminiPatchReview   & \CppGeminiPatchStyle \\
      & Claude-4.6-Sonnet & \CppClaudePatchBug   & \CppClaudePatchTest   & \CppClaudePatchReview   & \CppClaudePatchStyle \\
      & GPT-5-mini        & \CppGPTMiniPatchBug  & \CppGPTMiniPatchTest  & \CppGPTMiniPatchReview  & \CppGPTMiniPatchStyle \\
      & DeepSeek-V3.1     & \CppDeepSeekPatchBug & \CppDeepSeekPatchTest & \CppDeepSeekPatchReview & \CppDeepSeekPatchStyle \\
      & Qwen3-32B         & \CppQwenPatchBug     & \CppQwenPatchTest     & \CppQwenPatchReview     & \CppQwenPatchStyle \\
    \midrule
    \multirow{5}{*}{Java}   
      & Gemini-2.5-Flash  & \JavaGeminiPatchBug    & \JavaGeminiPatchTest    & \JavaGeminiPatchReview    & \JavaGeminiPatchStyle \\
      & Claude-4.6-Sonnet & \JavaClaudePatchBug    & \JavaClaudePatchTest    & \JavaClaudePatchReview    & \JavaClaudePatchStyle \\
      & GPT-5-mini        & \JavaGPTMiniPatchBug   & \JavaGPTMiniPatchTest   & \JavaGPTMiniPatchReview   & \JavaGPTMiniPatchStyle \\
      & DeepSeek-V3.1     & \JavaDeepSeekPatchBug  & \JavaDeepSeekPatchTest  & \JavaDeepSeekPatchReview  & \JavaDeepSeekPatchStyle \\
      & Qwen3-32B         & \JavaQwenPatchBug      & \JavaQwenPatchTest      & \JavaQwenPatchReview      & \JavaQwenPatchStyle \\
    \bottomrule
  \end{tabular}
\end{table*}

\begin{table*}[h]
\centering
\footnotesize
\caption{SWE-Agent performance on OmniCode across languages and models. These scores reflect the resolution rate of generated patches, disregarding instances where patches were not generated.}
\label{tab:sweagent_langs_capability}
\begin{tabular}{l l c c c c}
\toprule
\textbf{Language} & \textbf{Model} & \textbf{Bug-Fixing} & \textbf{Test-Generation} & \textbf{Review-Response} & \textbf{Style-Fixing} \\ 
\midrule
\multirow{5}{*}{Python}   
  & Gemini-2.5-Flash    &  40.6\%     & 15.2\%      & 32.3\%      & 62.2\% \\ 
  & Claude-4.6-Sonnet   & \textbf{73.2\%}      & \textbf{24.2}\%      & \textbf{83.1\%}      & \textbf{86.6\%} \\
  & GPT-5-mini          & 62.1\%     & 9.6\%     & 50.0\%     & 66.2\% \\
  & DeepSeek-V3.1       & 58.6\%    & 19.7\%    & 54.9\%    & 57.6\% \\  
  & Qwen3-32B           & 31.0\%     & 4.4\%     & 22.7\%        & 54.6\% \\ 
\midrule
\multirow{5}{*}{C++}    
  & Gemini-2.5-Flash    & 8.1\%     & 12.5\%     & 17.6\%     & 38.0\% \\
  & Claude-4.6-Sonnet   & 13.0\%     & \textbf{49.8\%}     & 11.1\%     & \textbf{100\%} \\
  & GPT-5-mini          & \textbf{24.3\%}    & 12.5\%    & 36.1\%    & 40.4\% \\
  & DeepSeek-V3.1       & 20.3\%   & 33.3\%   & \textbf{23.2\%}   & 21.9\% \\
  & Qwen3-32B           & 5.4\%       & 4.6\%       & 5.1\%       & 14.1\% \\
\midrule
\multirow{5}{*}{Java}   
  & Gemini-2.5-Flash    & 14.8\%    & 6.2\%    & 33.7\%    & 26.0\% \\
  & Claude-4.6-Sonnet   & \textbf{56.7\%}    & \textbf{43.8\%}    & \textbf{66.8}\%    & \textbf{36.4\%} \\
  & GPT-5-mini          & 47.9\%   & 5.9\%   & 51.3\%   & 44.2\% \\
  & DeepSeek-V3.1       & 33.3\%  & 24.0\%  & 48.6\%  & 25.4\% \\
  & Qwen3-32B           & 13.4\%      & 1.4\%      & 19.7\%      & 41.7\% \\
\bottomrule
\end{tabular}
\end{table*}

See Table~\ref{tab:sweagent_patch_generation_rate} and Table~\ref{tab:sweagent_langs_capability}.

\section{Failure Mode Analysis}
\begin{figure}[!htb]
\centering
    \includegraphics[scale=0.3]{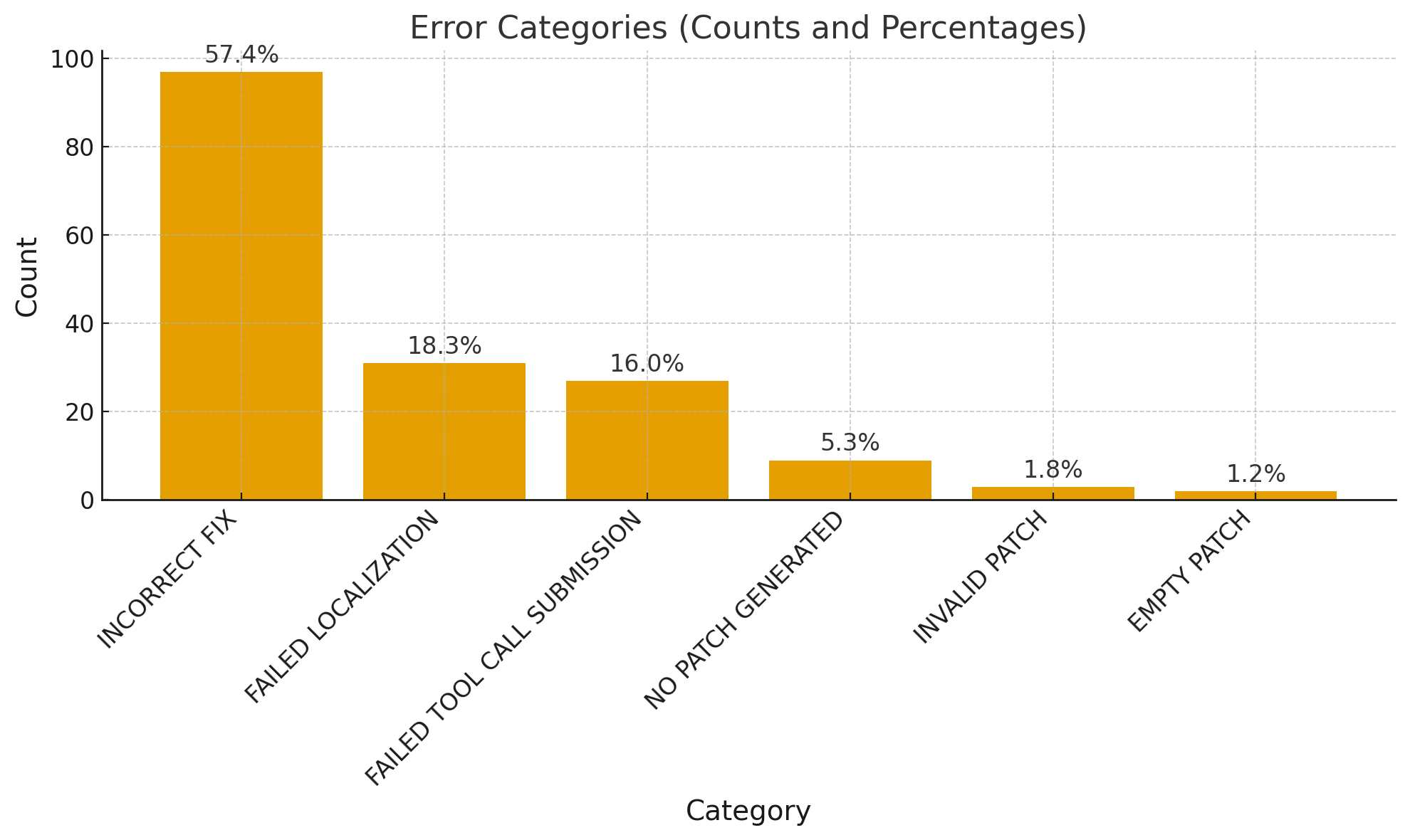}
    \caption{Distribution of agent failure modes for patches generated by SWE-Agent + Gemini-2.5-Flash on Bug-Fixing tasks.}
    \label{fig:failure-modes}
\end{figure}

\begin{figure*}
\centering
\includegraphics[width=\textwidth]{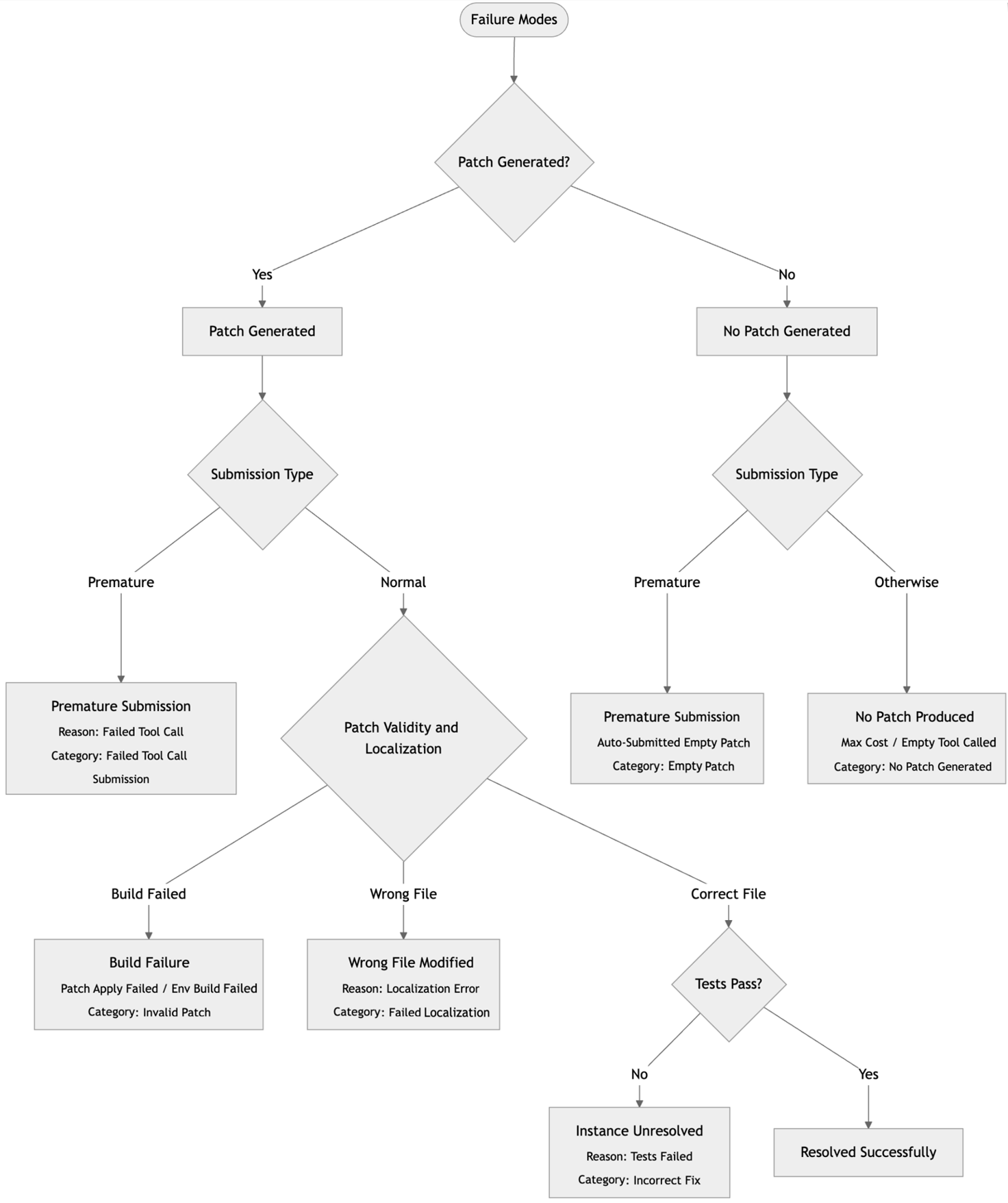}
\caption{Failure Mode Taxonomy of Agent-Generated Patches.}
\label{fig:failure-mode-taxonomy}
\end{figure*}

According to Figure~\ref{fig:failure-modes}, the most dominant failure mode is incorrect fixes, indicating that—even when an agent produces a patch—the modification is often semantically misaligned with the intended behavior. This highlights a core limitation in contemporary LLMs: difficulty performing deep program reasoning across multi-file, real-world repositories. Rather than syntactic invalidity, these failures reflect conceptual misunderstandings of logic, invariants, and dependencies.

The second major contributor is localization failure, suggesting that agents frequently misinterpret failing tests, select the wrong regions of code to modify, or struggle to trace error signals through complex call chains. 
While these results parallel observations in \cite{yang2025swesmithscalingdatasoftware}, we also quantify failed tool call submissions, which make up a significant portion of the errors.
These are cases where the agent outputs incorrect tool calls that trigger an auto-submission of the current patch it has.

The remaining categories represent smaller but meaningful sources of failure: no patch generated, invalid patches, and empty patches. These typically stem from toolchain integration issues or fragile execution pipelines, where agents fail to reliably produce well-formed edits. 

Taken together, these results suggest that the most effective pathways for improving SWE agents may need:
(1) strengthening semantic reasoning and behavioral alignment in code modifications,
(2) enhancing localization accuracy through richer retrieval and test-feedback interpretation, and
(3) developing more stable agentic control loops that prevent failed tool calls or premature termination and ensure patches undergo sufficient validation before submission.
We have included the methodology of our failure mode categorization in Figure~\ref{fig:failure-mode-taxonomy}.

\clearpage
\newpage
\section{Prompts}
\label{app:prompts}
\begin{tcblisting}{
  enhanced,
  breakable,
  title=Review Generation,
  listing only,
  listing options={
    basicstyle=\ttfamily,
    breaklines=true,
  }
}
You are an experienced software engineer tasked with reviewing code patches.
Below is a problem statement, a correct patch example, and a submitted patch that is likely incorrect or incomplete.
Please provide a detailed review of the submitted patch that identifies issues (e.g., missing context, incorrect modifications, or potential bugs) and specifies suggestions for improving the submitted patch so that it correctly solves the problem statement. 
Avoid referencing the correct patch directly.

Problem Statement:
{{ problem_statement }}

Correct Patch Example:
{{ correct_patch_example }}

Submitted Patch (Bad Patch):
{{ bad_patch }}

Detailed Review:

\end{tcblisting}

\begin{tcblisting}{
  enhanced,
  breakable,
  title=Bad Patch Generation,
  listing only,
  listing options={
    basicstyle=\ttfamily,
    breaklines=true,
  }
}
You are given a production-ready source file below. Your task:
1. **Introduce one to two subtle, functional bugs** without adding any comments
2. **Do NOT break compilation** and **do not introduce any syntax or spelling errors** or make any code-style changes.
3. **Do NOT change any import statements**
4. Preserve formatting and comments; modify only the minimum lines needed to trigger a logical failure under certain inputs.
5. Return **only** the full modified file content, with no explanations or diff markers.

--- {path} original content START ---
{curr_text}
--- {path} original content END ---

\end{tcblisting}

\begin{tcblisting}{
  enhanced,
  breakable,
  title=SWE-Agent Bug-fixing instructions,
  listing only,
  listing options={
    basicstyle=\ttfamily,
    breaklines=true,
  }
}
<uploaded_files>
{{working_dir}}
</uploaded_files>
I've uploaded a python code repository in the directory {{working_dir}}. Consider the following PR description:

<pr_description>
{{problem_statement}}
</pr_description>

Can you help me implement the necessary changes to the repository so that the requirements specified in the <pr_description> are met?
I've already taken care of all changes to any of the test files described in the <pr_description>. This means you DON'T have to modify the testing logic or any of the tests in any way!
Your task is to make the minimal changes to non-tests files in the {{working_dir}} directory to ensure the <pr_description> is satisfied.
Follow these steps to resolve the issue:
1. As a first step, it might be a good idea to find and read code relevant to the <pr_description>
2. Create a script to reproduce the error and execute it with `python <filename.py>` using the bash tool, to confirm the error
3. Edit the sourcecode of the repo to resolve the issue
4. Rerun your reproduce script and confirm that the error is fixed!
5. Think about edgecases and make sure your fix handles them as well
Your thinking should be thorough and so it's fine if it's very long.
\end{tcblisting}

\begin{tcblisting}{
  enhanced,
  breakable,
  title=SWE-Agent Test Generation instructions,
  listing only,
  listing options={
    basicstyle=\ttfamily,
    breaklines=true,
  }
}
<uploaded_files>
{{working_dir}}
</uploaded_files>
I've uploaded a python code repository in the directory {{working_dir}}. Consider the following problem description:

<problem_description>
{{problem_statement}}
</problem_description>

Can you help me implement a test that successfully reproduces the problem specified in the <problem_description>?
The test must be created in the repository's existing test suite and should be runnable with the repository's testing infrastructure / tooling (e.g. pytest).
Do not make any changes to the non-test code in the repository since we only need to create a reproduction test.
Follow these steps to resolve the issue:
1. As a first step, it might be a good idea to find and read code relevant to the <problem_description>
2. Create a script to reproduce the error and execute it with `python <filename.py>` using the bash tool, to confirm the error
3. Edit the the testing suite of the repo to implement a test based on this reproduction script which can be run using the repository's testing infrastructure / tooling (e.g. pytest)
4. Ensure this test runs and successfully reproduces the problem!
5. Remove the reproduction script and only keep changes to the test suite that reproduce the problem.
Your thinking should be thorough and so it's fine if it's very long.
\end{tcblisting}

\begin{tcblisting}{
  enhanced,
  breakable,
  title=SWE-Agent Style-Fix instructions,
  listing only,
  listing options={
    basicstyle=\ttfamily,
    breaklines=true,
  }
}
You have recently generated a patch to resolve an issue within this repository.
Pylint has been run on the modified files and has produced the following feedback:

{{problem_statement}}

Your task is to:
1. Analyze the Pylint violations provided in the problem statement
2. Understand the specific rules that were violated (e.g., naming conventions, unused imports, complexity issues)
3. Apply fixes that resolve these errors while maintaining code functionality
4. Ensure your changes follow Python best practices and improve code readability
5. Test that your fixes don't introduce new Pylint violations
6. Do not introduce any new files to fix the style errors

Common Pylint violations you may encounter:
- Naming and style issues (invalid-name, missing-docstring, line-too-long)
- Import issues (unused-import, wrong-import-order, reimported)
- Error-prone patterns (undefined-variable, no-member, unsubscriptable-object)
- Code design issues (too-many-arguments, too-many-locals, too-many-branches)
- Best practice and maintainability issues (fixme, unused-argument, broad-except)

Please resolve the Pylint feedback to the best of your ability, while preserving the functionality of the code.
Focus on the most critical violations first and ensure your fixes improve overall code quality and maintainability.
\end{tcblisting}

\begin{tcblisting}{
  enhanced,
  breakable,
  title=SWE-Agent Review-Fix instructions,
  listing only,
  listing options={
    basicstyle=\ttfamily,
    breaklines=true,
  }
}
<uploaded_files>
{{working_dir}}
</uploaded_files>
I've uploaded a code repository in the directory {{working_dir}}. {{problem_statement}}

Can you help me implement the necessary changes to the repository so that the requirements specified in the <pr_description> are met?
I've already taken care of all changes to any of the test files described in the <pr_description>. This means you DON'T have to modify the testing logic or any of the tests in any way!
Your task is to make the minimal changes to non-tests files in the {{working_dir}} directory to ensure the <pr_description> is satisfied.
Follow these steps to resolve the issue:
1. As a first step, it might be a good idea to find and read code relevant to the <pr_description>
2. Create a script to reproduce the error and execute it to confirm the error
3. Edit the sourcecode of the repo to resolve the issue
4. Rerun your reproduce script and confirm that the error is fixed!
5. Think about edgecases and make sure your fix handles them as well
Your thinking should be thorough and so it's fine if it's very long.
\end{tcblisting}

\subsection{Rulesets used for Style Review}\label{sec:style_rules_app}
See Table~\ref{tab:py-warnings}, Table~\ref{tab:java-warnings}, and Table~\ref{tab:cpp-warnings} for the list of style errors we account for in our Style Review tasks.

\begin{table*}[ht] \centering \scriptsize \caption{List of Python Style Errors.} \label{tab:py-warnings} \begin{tabularx}{\textwidth}{@{}XXX@{}} \toprule \texttt{\detokenize{protected-access}} & \texttt{\detokenize{redefined-outer-name}} & \texttt{\detokenize{unused-argument}} \\ \texttt{\detokenize{attribute-defined-outside-init}} & \texttt{\detokenize{abstract-method}} & \texttt{\detokenize{fixme}} \\ \texttt{\detokenize{redefined-builtin}} & \texttt{\detokenize{invalid-str-returned}} & \texttt{\detokenize{unused-variable}} \\ \texttt{\detokenize{anomalous-backslash-in-string}} & \texttt{\detokenize{unnecessary-pass}} & \texttt{\detokenize{broad-exception-caught}} \\ \texttt{\detokenize{raise-missing-from}} & \texttt{\detokenize{unbalanced-tuple-unpacking}} & \texttt{\detokenize{arguments-differ}} \\ \texttt{\detokenize{unused-import}} & \texttt{\detokenize{reimported}} & \texttt{\detokenize{assigning-non-slot}} \\ \texttt{\detokenize{unnecessary-lambda}} & \texttt{\detokenize{undefined-variable}} & \texttt{\detokenize{pointless-statement}} \\ \texttt{\detokenize{logging-fstring-interpolation}} & \texttt{\detokenize{missing-timeout}} & \texttt{\detokenize{unsubscriptable-object}} \\ \texttt{\detokenize{logging-not-lazy}} & \texttt{\detokenize{pointless-string-statement}} & \texttt{\detokenize{not-callable}} \\ \texttt{\detokenize{unspecified-encoding}} & \texttt{\detokenize{dangerous-default-value}} & \texttt{\detokenize{invalid-field-call}} \\ \texttt{\detokenize{possibly-used-before-assignment}} & \texttt{\detokenize{arguments-renamed}} & \texttt{\detokenize{eval-used}} \\ \texttt{\detokenize{no-self-argument}} & \texttt{\detokenize{unexpected-keyword-arg}} & \texttt{\detokenize{bare-except}} \\ \texttt{\detokenize{too-many-function-args}} & \texttt{\detokenize{no-value-for-parameter}} & \texttt{\detokenize{expression-not-assigned}} \\ \texttt{\detokenize{cell-var-from-loop}} & \texttt{\detokenize{comparison-with-callable}} & \texttt{\detokenize{super-init-not-called}} \\ \texttt{\detokenize{undefined-loop-variable}} & \texttt{\detokenize{used-before-assignment}} & \texttt{\detokenize{global-variable-not-assigned}} \\ \texttt{\detokenize{abstract-class-instantiated}} & \texttt{\detokenize{access-member-before-definition}} & \texttt{\detokenize{bad-staticmethod-argument}} \\ \texttt{\detokenize{deprecated-class}} & \texttt{\detokenize{function-redefined}} & \texttt{\detokenize{implicit-str-concat}} \\ \texttt{\detokenize{not-context-manager}} & \texttt{\detokenize{signature-differs}} & \texttt{\detokenize{super-without-brackets}} \\ \texttt{\detokenize{invalid-unary-operand-type}} & \texttt{\detokenize{broad-exception-raised}} & \texttt{\detokenize{arguments-out-of-order}} \\ \texttt{\detokenize{assert-on-string-literal}} & \texttt{\detokenize{bad-indentation}} & \texttt{\detokenize{global-statement}} \\ \texttt{\detokenize{global-variable-undefined}} & \texttt{\detokenize{import-self}} & \texttt{\detokenize{invalid-getnewargs-ex-returned}} \\ \texttt{\detokenize{invalid-metaclass}} & \texttt{\detokenize{invalid-repr-returned}} & \texttt{\detokenize{invalid-sequence-index}} \\ \texttt{\detokenize{isinstance-second-argument-not-valid-type}} & \texttt{\detokenize{keyword-arg-before-vararg}} & \texttt{\detokenize{misplaced-bare-raise}} \\ \texttt{\detokenize{missing-kwoa}} & \texttt{\detokenize{non-parent-init-called}} & \texttt{\detokenize{possibly-unused-variable}} \\ \texttt{\detokenize{raising-non-exception}} & \texttt{\detokenize{redundant-u-string-prefix}} & \texttt{\detokenize{redundant-unittest-assert}} \\ \texttt{\detokenize{subprocess-run-check}} & \texttt{\detokenize{unnecessary-ellipsis}} & \texttt{\detokenize{unused-private-member}} \\ \texttt{\detokenize{wildcard-import}} & \texttt{\detokenize{astroid-error}} & \texttt{\detokenize{syntax-error}} \\ \texttt{\detokenize{useless-parent-delegation}} & \texttt{\detokenize{bad-super-call}} & \texttt{\detokenize{method-hidden}} \\ \texttt{\detokenize{not-an-iterable}} & \texttt{\detokenize{too-few-format-args}} & \texttt{\detokenize{assignment-from-no-return}} \\ \texttt{\detokenize{assignment-from-none}} & \texttt{\detokenize{bad-chained-comparison}} & \texttt{\detokenize{bad-str-strip-call}} \\ \texttt{\detokenize{bad-string-format-type}} & \texttt{\detokenize{bad-thread-instantiation}} & \texttt{\detokenize{bidirectional-unicode}} \\ \texttt{\detokenize{contextmanager-generator-missing-cleanup}} & \texttt{\detokenize{deprecated-argument}} & \texttt{\detokenize{deprecated-method}} \\ \texttt{\detokenize{deprecated-module}} & \texttt{\detokenize{dict-iter-missing-items}} & \texttt{\detokenize{duplicate-except}} \\ \texttt{\detokenize{duplicate-key}} & \texttt{\detokenize{duplicate-string-formatting-argument}} & \texttt{\detokenize{duplicate-value}} \\ \texttt{\detokenize{exec-used}} & \texttt{\detokenize{f-string-without-interpolation}} & \texttt{\detokenize{format-string-without-interpolation}} \\ \texttt{\detokenize{inherit-non-class}} & \texttt{\detokenize{invalid-bool-returned}} & \texttt{\detokenize{invalid-length-returned}} \\ \texttt{\detokenize{invalid-overridden-method}} & \texttt{\detokenize{logging-format-interpolation}} & \texttt{\detokenize{logging-too-many-args}} \\ \texttt{\detokenize{lost-exception}} & \texttt{\detokenize{method-cache-max-size-none}} & \texttt{\detokenize{modified-iterating-list}} \\ \texttt{\detokenize{nested-min-max}} & \texttt{\detokenize{no-method-argument}} & \texttt{\detokenize{non-iterator-returned}} \\ \texttt{\detokenize{notimplemented-raised}} & \texttt{\detokenize{pointless-exception-statement}} & \texttt{\detokenize{positional-only-arguments-expected}} \\ \texttt{\detokenize{raising-bad-type}} & \texttt{\detokenize{raising-format-tuple}} & \texttt{\detokenize{redeclared-assigned-name}} \\ \texttt{\detokenize{redundant-keyword-arg}} & \texttt{\detokenize{return-in-finally}} & \texttt{\detokenize{return-in-init}} \\ \texttt{\detokenize{self-assigning-variable}} & \texttt{\detokenize{self-cls-assignment}} & \texttt{\detokenize{shadowed-import}} \\ \texttt{\detokenize{try-except-raise}} & \texttt{\detokenize{unbalanced-dict-unpacking}} & \texttt{\detokenize{undefined-all-variable}} \\ \texttt{\detokenize{unexpected-special-method-signature}} & \texttt{\detokenize{unnecessary-semicolon}} & \texttt{\detokenize{unpacking-non-sequence}} \\ \texttt{\detokenize{unreachable}} & \texttt{\detokenize{unsupported-assignment-operation}} & \texttt{\detokenize{unsupported-delete-operation}} \\ \texttt{\detokenize{unsupported-membership-test}} & \texttt{\detokenize{unused-format-string-argument}} & \texttt{\detokenize{unused-wildcard-import}} \\ \texttt{\detokenize{used-prior-global-declaration}} & \texttt{\detokenize{useless-else-on-loop}} & \texttt{\detokenize{using-constant-test}} \\ \bottomrule 
\end{tabularx} \end{table*}

\begin{table*}[htbp]
\centering
\scriptsize
\caption{List of Java Style Errors}
\label{tab:java-warnings}
\begin{tabularx}{\textwidth}{@{}XXX@{}}
\toprule
\texttt{\detokenize{AtLeastOneConstructor}} & \texttt{\detokenize{AvoidDuplicateLiterals}} & \texttt{\detokenize{CommentDefaultAccessModifier}} \\
\texttt{\detokenize{FieldNamingConventions}} & \texttt{\detokenize{LawOfDemeter}} & \texttt{\detokenize{LocalVariableCouldBeFinal}} \\
\texttt{\detokenize{MethodArgumentCouldBeFinal}} & \texttt{\detokenize{OnlyOneReturn}} & \texttt{\detokenize{ShortClassName}} \\
\texttt{\detokenize{UnnecessaryImport}} & \texttt{\detokenize{UseUtilityClass}} & \texttt{\detokenize{AvoidCatchingGenericException}} \\
\texttt{\detokenize{AvoidDeeplyNestedIfStmts}} & \texttt{\detokenize{AvoidLiteralsInIfCondition}} & \texttt{\detokenize{ClassWithOnlyPrivateConstructorsShouldBeFinal}} \\
\texttt{\detokenize{JUnitTestContainsTooManyAsserts}} & \texttt{\detokenize{JUnitTestsShouldIncludeAssert}} & \texttt{\detokenize{LinguisticNaming}} \\
\texttt{\detokenize{SystemPrintln}} & \texttt{\detokenize{TestClassWithoutTestCases}} & \texttt{\detokenize{AvoidAccessibilityAlteration}} \\
\texttt{\detokenize{AvoidCatchingThrowable}} & \texttt{\detokenize{CallSuperInConstructor}} & \texttt{\detokenize{CognitiveComplexity}} \\
\texttt{\detokenize{ImmutableField}} & \texttt{\detokenize{LooseCoupling}} & \texttt{\detokenize{ShortMethodName}} \\
\texttt{\detokenize{SignatureDeclareThrowsException}} & \texttt{\detokenize{TooManyStaticImports}} & \texttt{\detokenize{UseDiamondOperator}} \\
\texttt{\detokenize{UseUnderscoresInNumericLiterals}} & \texttt{\detokenize{UselessParentheses}} & \texttt{\detokenize{AssignmentInOperand}} \\
\texttt{\detokenize{AvoidFieldNameMatchingMethodName}} & \texttt{\detokenize{AvoidReassigningParameters}} & \texttt{\detokenize{AvoidThrowingRawExceptionTypes}} \\
\texttt{\detokenize{CollapsibleIfStatements}} & \texttt{\detokenize{ConfusingTernary}} & \texttt{\detokenize{CouplingBetweenObjects}} \\
\texttt{\detokenize{CyclomaticComplexity}} & \texttt{\detokenize{DataClass}} & \texttt{\detokenize{ExceptionAsFlowControl}} \\
\texttt{\detokenize{ExcessivePublicCount}} & \texttt{\detokenize{GodClass}} & \texttt{\detokenize{LiteralsFirstInComparisons}} \\
\texttt{\detokenize{MethodNamingConventions}} & \texttt{\detokenize{MutableStaticState}} & \texttt{\detokenize{NPathComplexity}} \\
\texttt{\detokenize{NcssCount}} & \texttt{\detokenize{NullAssignment}} & \texttt{\detokenize{PreserveStackTrace}} \\
\texttt{\detokenize{SimplifyBooleanReturns}} & \texttt{\detokenize{TooManyFields}} & \texttt{\detokenize{UnnecessaryBoxing}} \\
\texttt{\detokenize{UnnecessaryConstructor}} & \texttt{\detokenize{UnusedFormalParameter}} & \texttt{\detokenize{UseProperClassLoader}} \\
\texttt{\detokenize{AbstractClassWithoutAbstractMethod}} & \texttt{\detokenize{ArrayIsStoredDirectly}} & \texttt{\detokenize{AvoidBranchingStatementAsLastInLoop}} \\
\texttt{\detokenize{ClassNamingConventions}} & \texttt{\detokenize{CloseResource}} & \texttt{\detokenize{CompareObjectsWithEquals}} \\
\texttt{\detokenize{EmptyCatchBlock}} & \texttt{\detokenize{ExcessiveImports}} & \texttt{\detokenize{FieldDeclarationsShouldBeAtStartOfClass}} \\
\texttt{\detokenize{ForLoopCanBeForeach}} & \texttt{\detokenize{JUnit4TestShouldUseTestAnnotation}} & \texttt{\detokenize{LocalVariableNamingConventions}} \\
\texttt{\detokenize{OneDeclarationPerLine}} & \texttt{\detokenize{ReturnEmptyCollectionRatherThanNull}} & \texttt{\detokenize{UnnecessaryFullyQualifiedName}} \\
\texttt{\detokenize{UnnecessaryReturn}} & \texttt{\detokenize{UnnecessarySemicolon}} & \texttt{\detokenize{UnusedAssignment}} \\
\texttt{\detokenize{UseTryWithResources}} & \texttt{\detokenize{UseVarargs}} & \texttt{\detokenize{AvoidFieldNameMatchingTypeName}} \\
\texttt{\detokenize{AvoidReassigningLoopVariables}} & \texttt{\detokenize{AvoidUncheckedExceptionsInSignatures}} & \texttt{\detokenize{ControlStatementBraces}} \\
\texttt{\detokenize{EmptyControlStatement}} & \texttt{\detokenize{GenericsNaming}} & \texttt{\detokenize{GuardLogStatement}} \\
\texttt{\detokenize{MethodReturnsInternalArray}} & \texttt{\detokenize{PrematureDeclaration}} & \texttt{\detokenize{SwitchStmtsShouldHaveDefault}} \\
\texttt{\detokenize{UnnecessaryCast}} & \texttt{\detokenize{UnnecessaryModifier}} & \texttt{\detokenize{UnusedPrivateMethod}} \\
\texttt{\detokenize{UseLocaleWithCaseConversions}} & \texttt{\detokenize{UseShortArrayInitializer}} & \texttt{\detokenize{AvoidThrowingNullPointerException}} \\
\texttt{\detokenize{BooleanGetMethodName}} & \texttt{\detokenize{ConstantsInInterface}} & \texttt{\detokenize{ConstructorCallsOverridableMethod}} \\
\texttt{\detokenize{ExcessiveParameterList}} & \texttt{\detokenize{FinalFieldCouldBeStatic}} & \texttt{\detokenize{ForLoopVariableCount}} \\
\texttt{\detokenize{JUnitUseExpected}} & \texttt{\detokenize{MissingSerialVersionUID}} & \texttt{\detokenize{NonStaticInitializer}} \\
\texttt{\detokenize{OverrideBothEqualsAndHashcode}} & \texttt{\detokenize{UnnecessaryAnnotationValueElement}} & \texttt{\detokenize{UnnecessaryLocalBeforeReturn}} \\
\texttt{\detokenize{UnusedLocalVariable}} & \texttt{\detokenize{UseCollectionIsEmpty}} & \texttt{\detokenize{UseEqualsToCompareStrings}} \\
\texttt{\detokenize{UseStandardCharsets}} & \texttt{\detokenize{AbstractClassWithoutAnyMethod}} & \texttt{\detokenize{AvoidCatchingNPE}} \\
\texttt{\detokenize{AvoidProtectedFieldInFinalClass}} & \texttt{\detokenize{AvoidProtectedMethodInFinalClassNotExtending}} & \texttt{\detokenize{AvoidUsingHardCodedIP}} \\
\texttt{\detokenize{DoubleBraceInitialization}} & \texttt{\detokenize{EmptyMethodInAbstractClassShouldBeAbstract}} & \texttt{\detokenize{EqualsNull}} \\
\texttt{\detokenize{FormalParameterNamingConventions}} & \texttt{\detokenize{ImplicitSwitchFallThrough}} & \texttt{\detokenize{JUnit5TestShouldBePackagePrivate}} \\
\texttt{\detokenize{MissingStaticMethodInNonInstantiatableClass}} & \texttt{\detokenize{ReplaceVectorWithList}} & \texttt{\detokenize{SimpleDateFormatNeedsLocale}} \\
\texttt{\detokenize{SimplifiedTernary}} & \texttt{\detokenize{SwitchDensity}} & \texttt{\detokenize{AvoidDecimalLiteralsInBigDecimalConstructor}} \\
\texttt{\detokenize{AvoidDollarSigns}} & \texttt{\detokenize{AvoidInstanceofChecksInCatchClause}} & \texttt{\detokenize{AvoidPrintStackTrace}} \\
\texttt{\detokenize{AvoidRethrowingException}} & \texttt{\detokenize{AvoidStringBufferField}} & \texttt{\detokenize{DoNotCallGarbageCollectionExplicitly}} \\
\texttt{\detokenize{DontImportSun}} & \texttt{\detokenize{FinalParameterInAbstractMethod}} & \texttt{\detokenize{IdenticalCatchBranches}} \\
\texttt{\detokenize{MissingOverride}} & \texttt{\detokenize{NonSerializableClass}} & \texttt{\detokenize{PrimitiveWrapperInstantiation}} \\
\texttt{\detokenize{SuspiciousEqualsMethodName}} & \texttt{\detokenize{UnusedPrivateField}} & \texttt{\detokenize{AvoidThrowingNewInstanceOfSameException}} \\
\texttt{\detokenize{DefaultLabelNotLastInSwitchStmt}} & \texttt{\detokenize{DetachedTestCase}} & \texttt{\detokenize{DoNotExtendJavaLangThrowable}} \\
\texttt{\detokenize{DoNotTerminateVM}} & \texttt{\detokenize{ForLoopShouldBeWhileLoop}} & \texttt{\detokenize{InstantiationToGetClass}} \\
\texttt{\detokenize{JUnit4SuitesShouldUseSuiteAnnotation}} & \texttt{\detokenize{JumbledIncrementer}} & \texttt{\detokenize{LogicInversion}} \\
\texttt{\detokenize{ProperCloneImplementation}} & \texttt{\detokenize{ReplaceHashtableWithMap}} & \texttt{\detokenize{SimplifyBooleanExpressions}} \\
\texttt{\detokenize{SimplifyConditional}} & \texttt{\detokenize{SingletonClassReturningNewInstance}} & \texttt{\detokenize{SingularField}} \\
\texttt{\detokenize{UseObjectForClearerAPI}} & \texttt{\detokenize{AssignmentToNonFinalStatic}} & \texttt{\detokenize{AvoidMessageDigestField}} \\
\texttt{\detokenize{AvoidMultipleUnaryOperators}} & \texttt{\detokenize{AvoidUsingOctalValues}} & \texttt{\detokenize{CheckSkipResult}} \\
\texttt{\detokenize{ClassCastExceptionWithToArray}} & \texttt{\detokenize{CloneMethodMustBePublic}} & \texttt{\detokenize{CloneMethodMustImplementCloneable}} \\
\texttt{\detokenize{CloneMethodReturnTypeMustMatchClassName}} & \texttt{\detokenize{DoNotExtendJavaLangError}} & \texttt{\detokenize{DoNotHardCodeSDCard}} \\
\texttt{\detokenize{DoNotThrowExceptionInFinally}} & \texttt{\detokenize{DontUseFloatTypeForLoopIndices}} & \texttt{\detokenize{FinalizeDoesNotCallSuperFinalize}} \\
\texttt{\detokenize{InvalidLogMessageFormat}} & \texttt{\detokenize{NoPackage}} & \texttt{\detokenize{PackageCase}} \\
\texttt{\detokenize{SingleMethodSingleton}} & \texttt{\detokenize{UnconditionalIfStatement}} & \texttt{\detokenize{UnnecessaryCaseChange}} \\
\texttt{\detokenize{UnusedNullCheckInEquals}} & \texttt{\detokenize{UseExplicitTypes}} & \texttt{\detokenize{UselessOperationOnImmutable}} \\
\texttt{\detokenize{UselessOverridingMethod}} & \texttt{\detokenize{UselessQualifiedThis}} & \texttt{\detokenize{WhileLoopWithLiteralBoolean}} \\
\bottomrule
\end{tabularx}
\end{table*}

\begin{table*}[htbp]
\centering
\scriptsize
\caption{List of CPP Style Errors}
\label{tab:cpp-warnings}
\begin{tabularx}{\textwidth}{@{}XXX@{}}
\toprule
\texttt{\detokenize{misc-include-cleaner}} & \texttt{\detokenize{misc-use-anonymous-namespace}} \\
\texttt{\detokenize{cppcoreguidelines-avoid-magic-numbers}} & \texttt{\detokenize{cppcoreguidelines-avoid-do-while}} \\
\texttt{\detokenize{misc-const-correctness}} & \texttt{\detokenize{cppcoreguidelines-rvalue-reference-param-not-moved}} \\
\texttt{\detokenize{misc-non-private-member-variables-in-classes}} & \texttt{\detokenize{bugprone-easily-swappable-parameters}} \\
\texttt{\detokenize{cppcoreguidelines-pro-bounds-pointer-arithmetic}} & \texttt{\detokenize{cppcoreguidelines-avoid-c-arrays}} \\
\texttt{\detokenize{cppcoreguidelines-avoid-non-const-global-variables}} & \texttt{\detokenize{cppcoreguidelines-pro-bounds-array-to-pointer-decay}} \\
\texttt{\detokenize{cppcoreguidelines-owning-memory}} & \texttt{\detokenize{cppcoreguidelines-init-variables}} \\
\texttt{\detokenize{cppcoreguidelines-macro-usage}} & \texttt{\detokenize{cppcoreguidelines-special-member-functions}} \\
\texttt{\detokenize{cppcoreguidelines-pro-type-member-init}} & \texttt{\detokenize{cppcoreguidelines-pro-type-static-cast-downcast}} \\
\texttt{\detokenize{misc-no-recursion}} & \texttt{\detokenize{performance-enum-size}} \\
\texttt{\detokenize{bugprone-narrowing-conversions}} & \texttt{\detokenize{cppcoreguidelines-narrowing-conversions}} \\
\texttt{\detokenize{cppcoreguidelines-pro-type-reinterpret-cast}} & \texttt{\detokenize{cppcoreguidelines-pro-type-union-access}} \\
\texttt{\detokenize{cppcoreguidelines-use-default-member-init}} & \texttt{\detokenize{cppcoreguidelines-pro-bounds-constant-array-index}} \\
\texttt{\detokenize{bugprone-implicit-widening-of-multiplication-result}} & \texttt{\detokenize{bugprone-macro-repeated-side-effects}} \\
\texttt{\detokenize{bugprone-suspicious-include}} & \texttt{\detokenize{clang-analyzer-optin.core.EnumCastOutOfRange}} \\
\texttt{\detokenize{cppcoreguidelines-avoid-const-or-ref-data-members}} & \texttt{\detokenize{cppcoreguidelines-explicit-virtual-functions}} \\
\texttt{\detokenize{cppcoreguidelines-pro-type-vararg}} & \texttt{\detokenize{portability-simd-intrinsics}} \\
\bottomrule
\end{tabularx}
\end{table*}

% \clearpage

\subsection{Style Review Error Analysis}
\label{style_review_analysis}

\begin{figure*}
\centering
\includegraphics[scale=0.4]{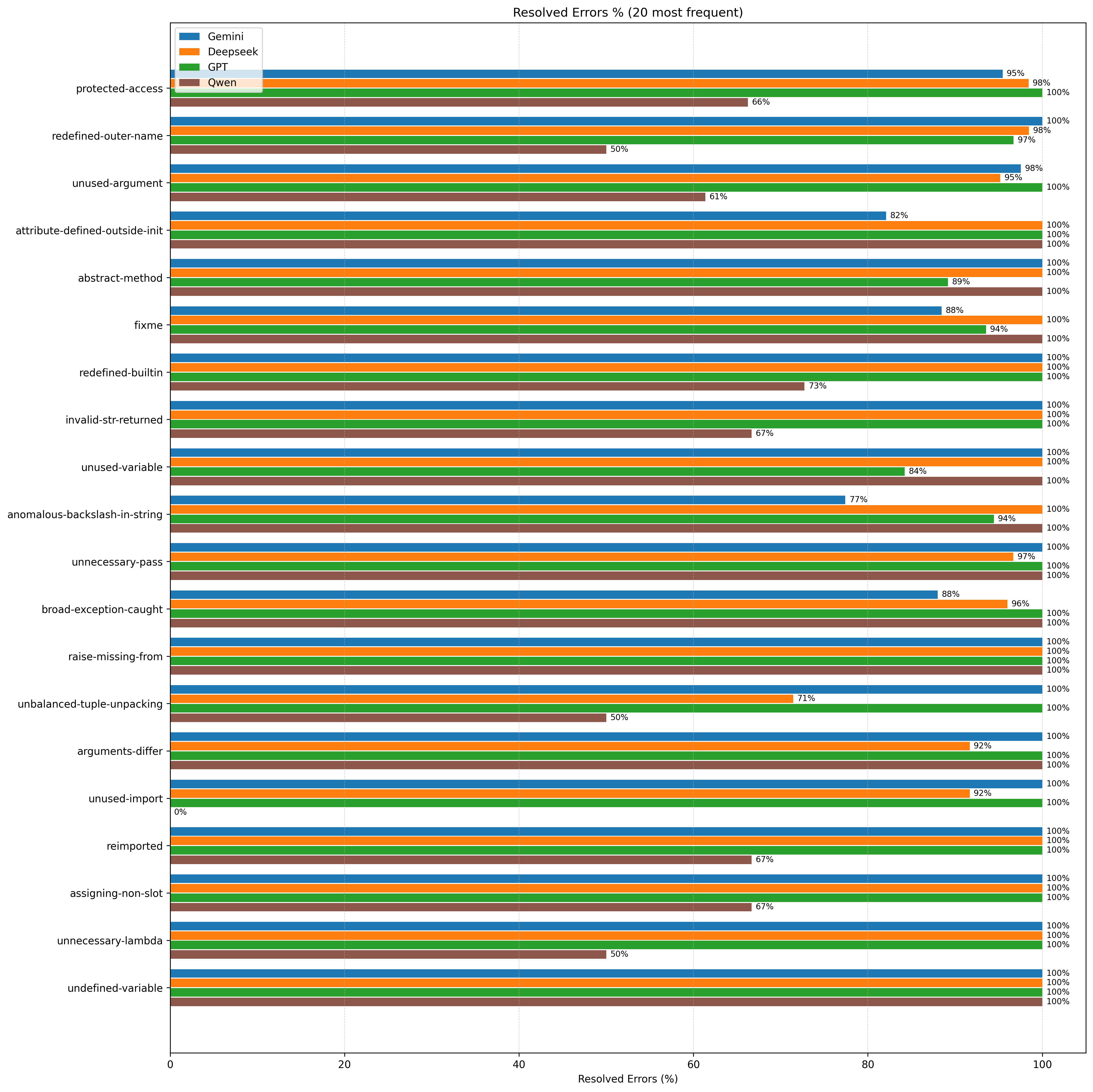}
\caption{Resolve rates for the 20 most frequent style errors in Python.}
\label{fig:resolve-rate-python}
\end{figure*}

In Figure~\ref{fig:resolve-rate-python}, it can be seen that Gemini consistently fixes structural and naming issues, such as redefined names, protected access, and undefined variables, which reduces runtime errors and improves code stability. Deepseek is strongest on module and resource patterns such as unused imports, reimported modules, and unnecessary lambdas, producing leaner code and fewer maintenance surprises. GPT performs best on interface and initialization problems like unused arguments, mismatched function signatures, and attribute initialization, lowering the risk of subtle API misuse. Qwen provides broad, reliable cleanups across stylistic and warning classes, including broad exception handling and anomalous string escapes, raising overall code quality and guideline compliance.

\begin{figure*}
\centering
\includegraphics[scale=0.4]{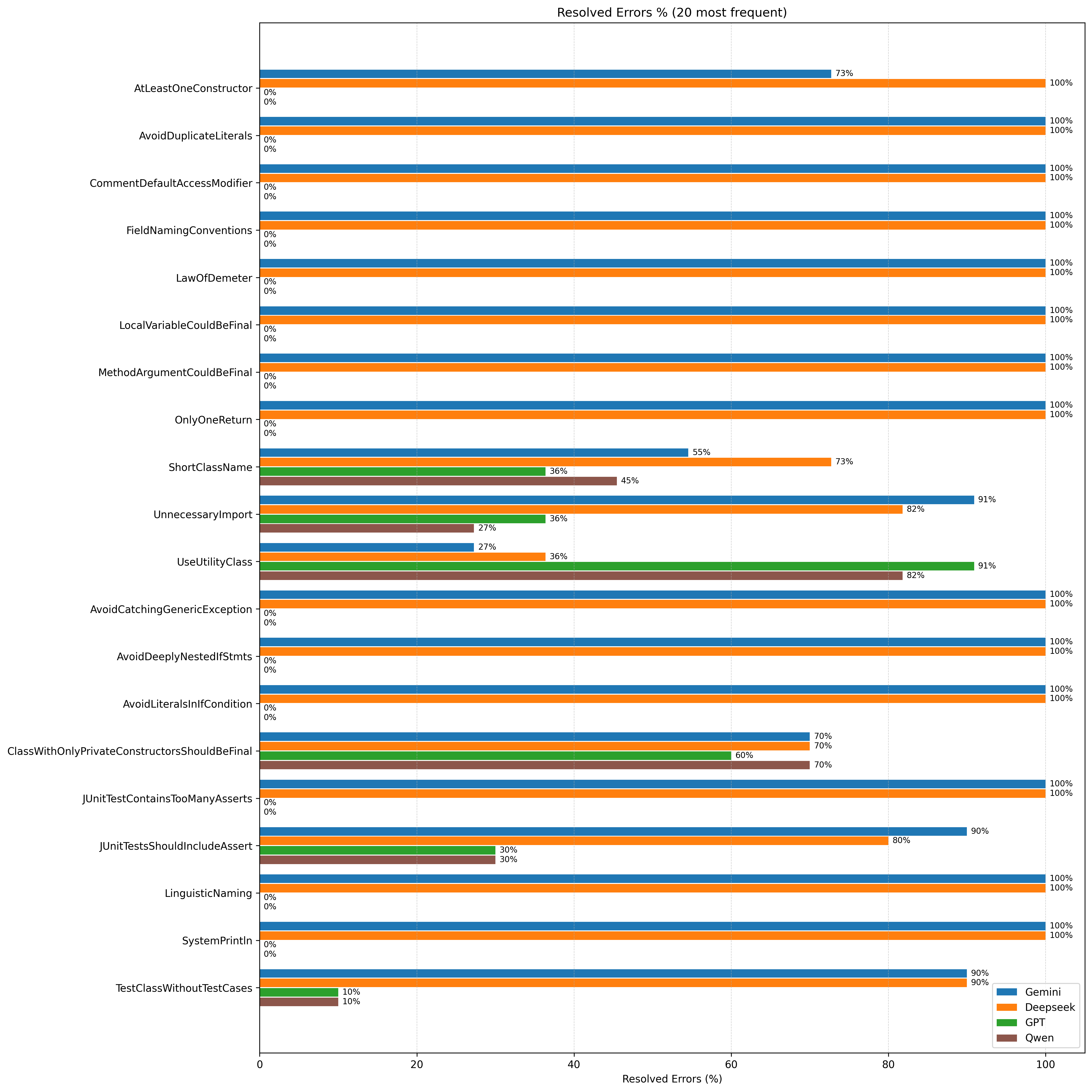}
\caption{Resolve rates for the 20 most frequent style errors in Java.}
\label{fig:resolve-rate-java}
\end{figure*}

Figure~\ref{fig:resolve-rate-java} shows that Gemini is strongest at structural and object-oriented fixes, such as adding or normalizing constructors, enforcing field and method naming conventions, and flattening deeply nested logic, which improves maintainability and reduces subtle runtime defects. Deepseek performs best on cleanup and refactor tasks like removing unnecessary imports, eliminating duplicate literals, and simplifying utility classes, which reduces code bloat and lowers maintenance costs. GPT excels at test and API related fixes, including adding assertions, finalizing method arguments and local variables, and correcting single return patterns, which raises test reliability and prevents interface regressions. Qwen provides consistent, broad stylistic cleanups such as enforcing short class name policies and small local correctness adjustments.

\begin{figure*}
\centering
\includegraphics[scale=0.4]{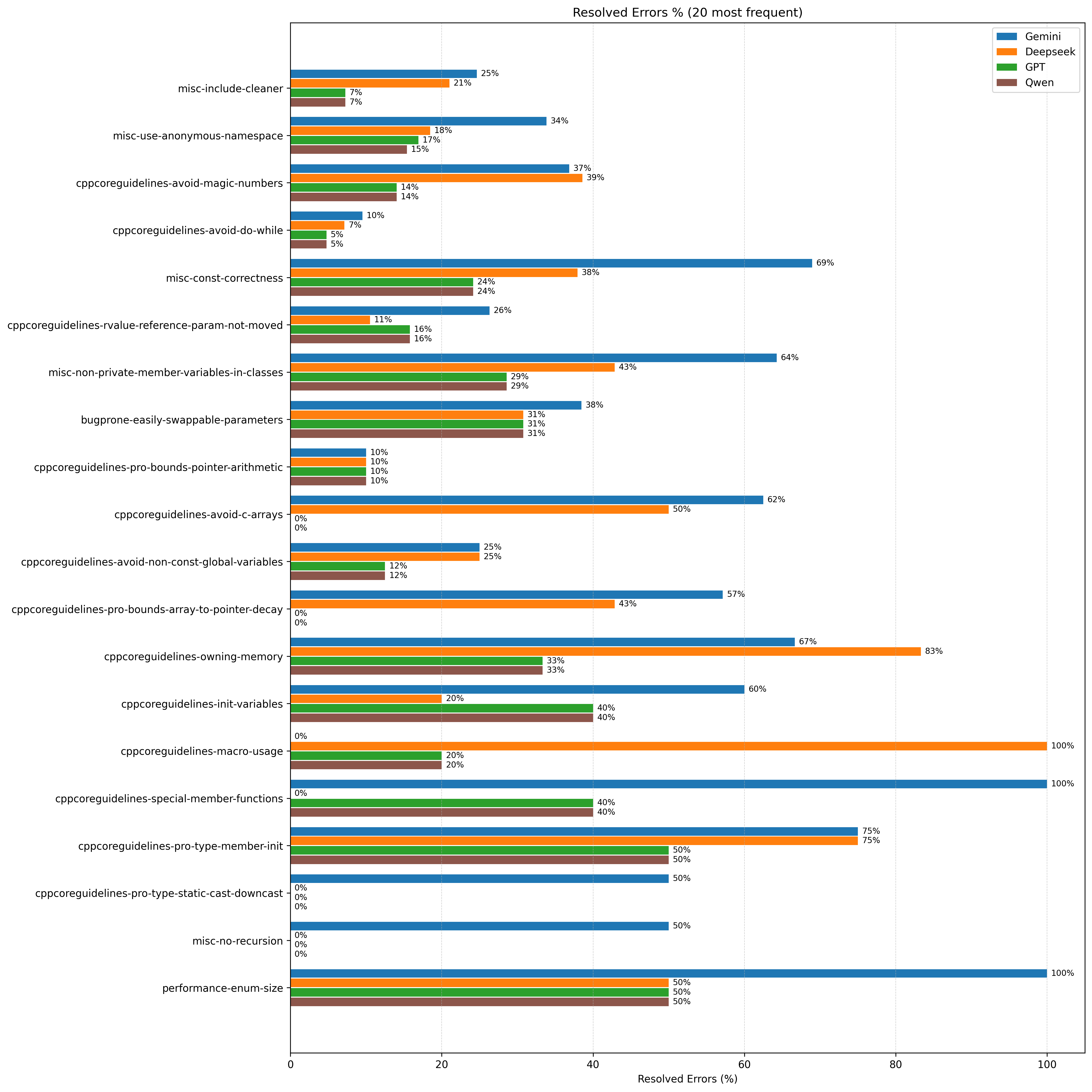}
\caption{Resolve rates for the 20 most frequent style errors in CPP.}
\label{fig:resolve-rate-cpp}
\end{figure*}

Figure~\ref{fig:resolve-rate-cpp} shows that Gemini excels at correctness and structural issues such as constant correctness, non-private member variables, and avoiding C-style arrays; fixing these issues reduces undefined behavior and lowers crash risk. Deepseek is strongest on memory management and macro usage, addressing owning memory and macro patterns that directly improve resource safety and long term maintainability. GPT performs best at initialization and API correctness tasks, such as initializing variables and easily swappable parameters, which prevents subtle bugs and makes interfaces safer to use. Qwen delivers steady, moderate improvements across stylistic and guideline rules, making it a good choice for broad cleanup and incremental rule conformance. 

Through these insights, we observe that LLMs excel at deterministic, local rewrites (e.g., removing duplicate imports, adding simple initializers, replacing macros with constexpr, extracting constants, tightening exception catches, and adding final) because they preserve observable behavior in the common case. They falter on fixes that require global program understanding, API-level decisions, ownership/lifetime tradeoffs, or knowledge about intended public interfaces, exactly the places where a human's design judgment, test coverage, and knowledge of downstream users matter.

\begin{comment}
\begin{figure}
\centering
\includegraphics[scale=0.4]{java_additional_errors.png}
\caption{Counts for the 20 most frequent additional errors in Java.}
\label{fig:additional-errors-java}
\end{figure}
\end{comment}

\section{Correlation Analysis across Tasks} 
\label{sec:task_correlation}

\begin{figure*}
\centering
\includegraphics[scale=0.4] {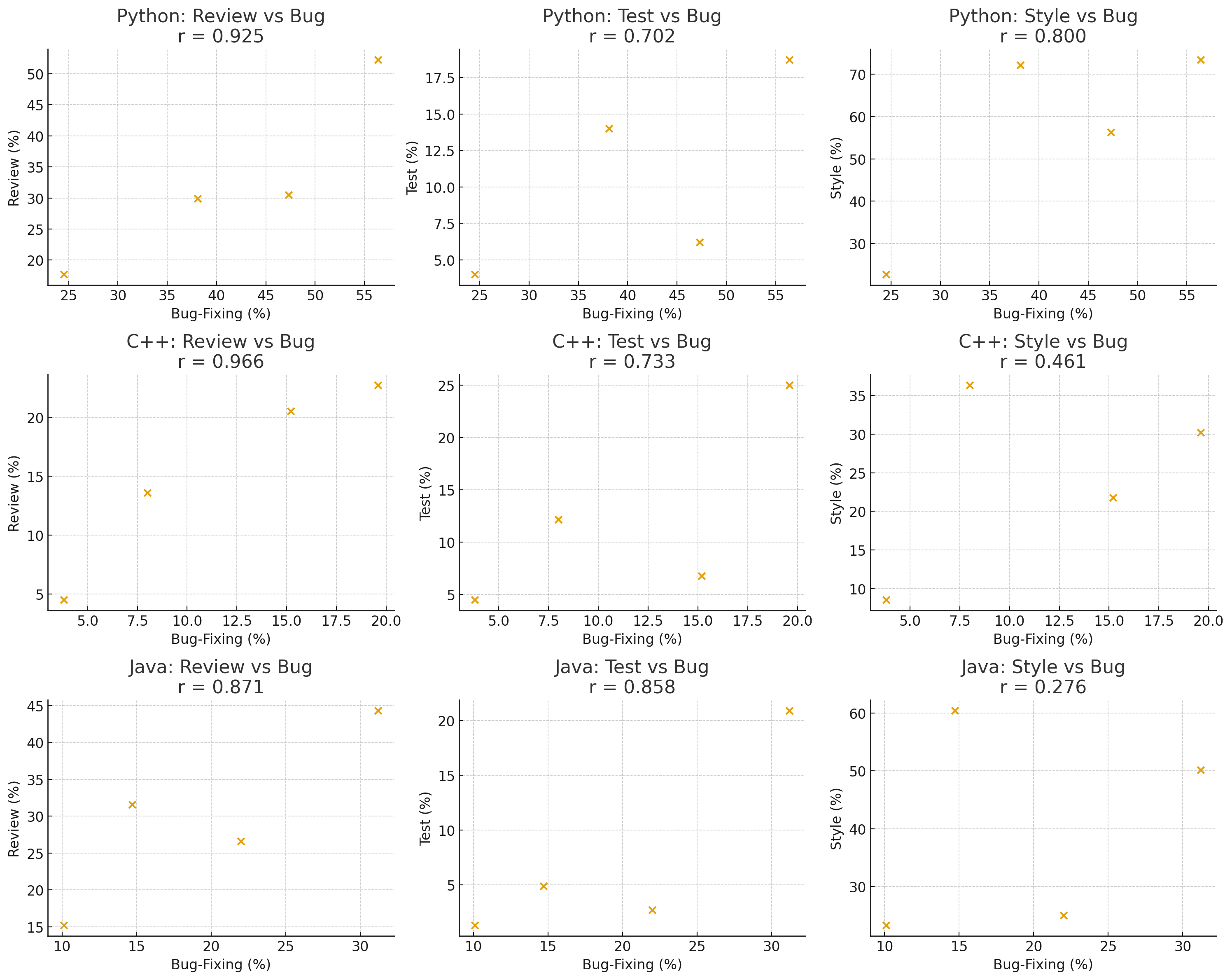}
    \caption{Bugfixing performance plotted together with performance on other tasks separately for each language.}
    \label{fig:correlation_analysis}
\end{figure*}

See Figure~\ref{fig:correlation_analysis}.

\begin{table*}[h]
  \centering
  \footnotesize % Use smaller font size as in the example
  \caption{Review-Response - Avg. Complexity Score.}
  \label{table:reviewfix_complexity}
  
  % New table structure with 8 columns, using \multirow
  \begin{tabular}{l l cccccc}
    \toprule % Top line from booktabs
    % Bold headers, with stacking for long ones
    \textbf{Model} & \textbf{Language} & \multicolumn{1}{c}{\textbf{\begin{tabular}[c]{@{}c@{}}Gold Avg\\Complexity\end{tabular}}} & \textbf{\begin{tabular}[c]{@{}c@{}}Model Avg\\Complexity\end{tabular}} & \multicolumn{1}{c}{\textbf{\begin{tabular}[c]{@{}c@{}}Resolved\\Gold Avg\\Complexity\end{tabular}}} & \multicolumn{1}{c}{\textbf{\begin{tabular}[c]{@{}c@{}}Resolved\\Model Avg\\Complexity\end{tabular}}} & \multicolumn{1}{c}{\textbf{\begin{tabular}[c]{@{}c@{}}Unresolved\\Gold Avg\\Complexity\end{tabular}}} & \multicolumn{1}{c}{\textbf{\begin{tabular}[c]{@{}c@{}}Unresolved\\Model Avg\\Complexity\end{tabular}}} \\
    \midrule
    \multirow{3}{*}{Gemini 2.5 flash} & Python & 7.07 & 1635.47 & 3.69 & 9.58 & 7.93 & 2408.95 \\
    & Java & 19.24 & 9.95 & 6.49 & 6.86 & 17.25 & 11.53 \\
    & C++ & 47.55 & 128.33 & 5.98 & 4.85 & 41.85 & 154.79 \\
    \midrule
    \multirow{3}{*}{GPT-5-mini} & Python & 7.07 & 289.22 & 4.96 & 13.80 & 7.40 & 543.45 \\
    & Java & 19.24 & 6.26 & 6.91 & 5.58 & 15.32 & 6.99 \\
    & C++ & 47.55 & 955.24 & 9.01 & 5.84 & 12.39 & 1489.28 \\
    \midrule
    \multirow{3}{*}{Deepseek v3.1} & Python & 7.07 & 10.71 & 4.36 & 6.24 & 9.23 & 16.26 \\
    & Java & 19.24 & 9.75 & 6.69 & 7.04 & 19.67 & 12.32 \\
    & C++ & 47.55 & 195.10 & 8.07 & 6.28 & 38.26 & 252.31 \\
    \midrule
    \multirow{3}{*}{qwen3-32b} & Python & 7.07 & 519.86 & 3.15 & 3.26 & 7.41 & 639.08 \\
    & Java & 19.24 & 3.96 & 5.85 & 2.83 & 16.83 & 4.31 \\
    & C++ & 47.55 & 248.65 & 2.25 & 2.4 & 14.55 & 261.96 \\
    \bottomrule
  \end{tabular}
\end{table*}

\section{Details of Data Collection}
\label{app:data}

% \Fix{Add numbers of added instances for Python and C++}
We consider popular projects, filtering out tutorials and other non-code repositories.
From these, we collect merged pull requests that (1) resolve an issue and (2) introduce at least one test.
To ensure that each instance is a meaningful task for an agent to be evaluated on, we perform manual inspection. 
Only instances where the changes introduced in the pull request are within the scope of the description of the issue are kept.
For reproducibility, we also filter out flaky and non-functional instances where dependencies could not be installed properly or where pre-existing tests would not pass consistently. 
In addition, we also discard issues if they only involve simpler changes to documentation or configuration files.

\section{Details of Bad Patch Generation Failures}
\label{app:badpatchgenstats}

We use two approaches for bad patch generation as detailed in Section~\ref{sec:testgen}: (1) collecting failed bug-fixing attempts and (2) perturbing the correct patch to add bugs.
For Python, we aimed for three patches per instance using approach 1. Most Python instances have all three, but around 1/3 don't have an approach 1 patch.
All Python instances have one patch using approach 2.
For Java, all instances have three approaches, using approach 1.
For C++, most instances have 3 patches using approach 1.
Overall, there are 112 bad patches for C++, 237 for Java, and 760 (487 from approach 1; 273 from approach 2) for Python.

\section{Human Evaluation of Synthetic Data}
\label{app:badpatchhumaneval}

As we rely on synthetically generated patches in our work,  human evaluation is an important initial validation step. For this purpose, six members of our research team independently inspected a subset of reviews and bad patches, verifying that (1) reviews provide meaningful guidance without leaking the gold patch, and (2) bad patches are substantively incorrect.

The questions asked for each category were - 

\begin{itemize}
\item Patch Incorrectness Type: Why is the bad patch incorrect?
\item Patch Plausibility: How convincing does the bad patch look?
\item Review Information Leakage: Does the review reveal the correct fix?
\item Review Correctness: Does the review accurately identify what's wrong?
 
\end{itemize}

Table~\ref{tab:human_eval} summarizes the aggregated results. Most bad patches reflect wrong but plausible approaches rather than trivial mistakes. Specifically, 51.1\% of bad patches follow an incorrect approach, while only 14.4\% are trivially wrong. On a 1--3 plausibility scale (1 = obviously wrong, 2 = somewhat plausible, 3 = very convincing), the mean plausibility score is 2.10, indicating that most bad patches are at least somewhat plausible, with relatively few being either trivially wrong or highly convincing.

With regard to review quality, reviews rarely leak the solution directly: 70.0\% exhibit no leakage, 30.0\% exhibit moderate leakage, and none exhibit severe leakage. In addition, 88.9\% of reviews are fully correct and another 10.0\% are partially correct, meaning that 98.9\% are at least partially correct overall. Inter-annotator agreement is moderate to good, with majority agreement in at least 86.7\% of instances across all categories.

Overall, these results suggest that the generated tasks are neither trivial nor degenerate, and that reviews generally provide constructive, non-revealing feedback. Together with the LLM-based categorization, this provides an initial validation of the OmniCode task generation pipeline and its diversity.

\begin{table}[t]
    \centering
    \caption{Human evaluation (n=6) results for synthetic reviews and bad patches across 20 task instances.}
    \label{tab:human_eval}
    \begin{tabular}{p{0.26\columnwidth} p{0.42\columnwidth} p{0.18\columnwidth}}
        \toprule
        \textbf{Category} & \textbf{Subcategory} & \textbf{Score} \\
        \midrule
        \multirow{4}{=}{\raggedright Patch Incorrectness Type}
            & Trivially Wrong & 14.4\% \\
            & Wrong Approach & 51.1\% \\
            & Gold + Bugs & 12.2\% \\
            & Incomplete Fix & 22.2\% \\
        \midrule
        Patch Plausibility & Mean (1--3) & 2.10 \\
        \midrule
        \multirow{3}{=}{\raggedright Review Information Leakage}
            & None & 70.0\% \\
            & Moderate & 30.0\% \\
            & Severe & 0.0\% \\
        \midrule
        \multirow{3}{=}{\raggedright Review Correctness}
            & Correct & 88.9\% \\
            & Partially Correct & 10.0\% \\
            & Incorrect & 1.1\% \\
        \bottomrule
    \end{tabular}
\end{table}

\section{Confidence Intervals}
\label{app:confidence_intervals}
In addition to our performance evaluation, we compute confidence intervals on the main metrics presented in the paper in Tables~\ref{tab:confidence_intervals_main} and~\ref{tab:confidence_intervals_agents}.

\begin{table*}[h]
  \centering
  \footnotesize
  \caption{Confidence intervals for the main evaluation metrics across models and tasks. Values are reported as a percentage $\pm$ confidence interval.}
  \label{tab:confidence_intervals_main}
  \begin{tabular}{l l c c c c}
    \toprule
    \textbf{Language} & \textbf{Model} & \textbf{Bug-Fixing} & \textbf{Test-Generation} & \textbf{Review-Response} & \textbf{Style-Fixing} \\
    \midrule
    \multirow{4}{*}{Python}
      & Gemini-2.5-Flash & 38.1\% $\pm$ 5.8 & 14.0\% $\pm$ 5.3 & 29.9\% $\pm$ 7.0 & 57.0\% $\pm$ 5.8 \\
      & GPT-5-mini       & 47.3\% $\pm$ 5.9 & 6.2\% $\pm$ 3.7  & 30.5\% $\pm$ 7.1 & 45.9\% $\pm$ 6.6 \\
      & DeepSeek-V3.1    & 56.4\% $\pm$ 5.9 & 18.7\% $\pm$ 6.0 & 52.2\% $\pm$ 7.7 & 54.0\% $\pm$ 5.3 \\
      & Qwen3-32B        & 24.5\% $\pm$ 5.1 & 4.0\% $\pm$ 3.0  & 17.7\% $\pm$ 5.8 & 19.5\% $\pm$ 5.8 \\
    \midrule
    \multirow{4}{*}{C++}
      & Gemini-2.5-Flash & 8.0\% $\pm$ 5.0  & 12.2\% $\pm$ 9.7 & 13.6\% $\pm$ 10.1 & 30.5\% $\pm$ 5.4 \\
      & GPT-5-mini       & 15.2\% $\pm$ 6.7 & 6.8\% $\pm$ 7.4  & 20.5\% $\pm$ 11.9 & 19.5\% $\pm$ 5.5 \\
      & DeepSeek-V3.1    & 19.6\% $\pm$ 7.4 & 25.0\% $\pm$ 12.8 & 22.7\% $\pm$ 12.4 & 18.8\% $\pm$ 3.7 \\
      & Qwen3-32B        & 3.8\% $\pm$ 3.5  & 4.5\% $\pm$ 6.1  & 4.5\% $\pm$ 6.1  & 6.7\% $\pm$ 3.0 \\
    \midrule
    \multirow{4}{*}{Java}
      & Gemini-2.5-Flash & 14.7\% $\pm$ 6.7 & 4.9\% $\pm$ 4.8  & 31.6\% $\pm$ 10.3 & 22.0\% $\pm$ 2.4 \\
      & GPT-5-mini       & 22.0\% $\pm$ 7.8 & 2.7\% $\pm$ 3.6  & 26.6\% $\pm$ 9.7  & 22.1\% $\pm$ 5.6 \\
      & DeepSeek-V3.1    & 31.2\% $\pm$ 8.7 & 20.9\% $\pm$ 9.0 & 44.3\% $\pm$ 10.9 & 23.1\% $\pm$ 2.9 \\
      & Qwen3-32B        & 10.1\% $\pm$ 5.7 & 1.3\% $\pm$ 2.5  & 15.2\% $\pm$ 7.9  & 18.5\% $\pm$ 4.8 \\
    \bottomrule
  \end{tabular}
\end{table*}

\begin{table*}[h]
  \centering
  \footnotesize
  \caption{Confidence intervals for SWE-Agent and Aider comparison across tasks. Values are reported as percentage $\pm$ confidence interval.}
  \label{tab:confidence_intervals_agents}
  \begin{tabular}{l l c c c c}
    \toprule
    \textbf{Language} & \textbf{Agent} & \textbf{Bug-Fixing} & \textbf{Test-Generation} & \textbf{Review-Response} & \textbf{Style-Fixing} \\
    \midrule
    \multirow{2}{*}{Python}
      & SWE-Agent & 38.1\% $\pm$ 5.8 & 14.0\% $\pm$ 5.3 & 29.9\% $\pm$ 7.0 & 57.0\% $\pm$ 5.8 \\
      & Aider     & 32.4\% $\pm$ 5.6 & 9.4\% $\pm$ 4.5  & 26.8\% $\pm$ 6.8 & 48.6\% $\pm$ 6.4 \\
    \midrule
    \multirow{2}{*}{C++}
      & SWE-Agent & 8.0\% $\pm$ 5.0  & 12.2\% $\pm$ 9.7 & 13.6\% $\pm$ 10.1 & 30.5\% $\pm$ 5.4 \\
      & Aider     & 1.8\% $\pm$ 2.5  & 2.3\% $\pm$ 4.4  & 4.5\% $\pm$ 6.1   & 7.9\% $\pm$ 3.8 \\
    \midrule
    \multirow{2}{*}{Java}
      & SWE-Agent & 14.7\% $\pm$ 6.7 & 4.9\% $\pm$ 4.8  & 31.6\% $\pm$ 10.3 & 22.0\% $\pm$ 2.4 \\
      & Aider     & 19.3\% $\pm$ 7.4 & 3.9\% $\pm$ 9.6  & 25.3\% $\pm$ 9.6  & 21.7\% $\pm$ 3.1 \\
    \bottomrule
  \end{tabular}
\end{table*}

\end{document}

%% file: math_commands.tex
\usepackage{amsmath,amsfonts,bm}

\def\eqref#1{equation~\ref{#1}}
\def\1{\bm{1}}

\DeclareMathAlphabet{\mathsfit}{\encodingdefault}{\sfdefault}{m}{sl}
\SetMathAlphabet{\mathsfit}{bold}{\encodingdefault}{\sfdefault}{bx}{n}

%% file: macros.tex
\newcommand{\tool}{OmniCode\xspace}

\newcommand{\sweagent}{SWE-Agent\xspace}
\newcommand{\aider}{Aider\xspace}
\newcommand{\gemini}{Gemini-2.5-Flash\xspace}
\newcommand{\gptmini}{GPT-5-mini\xspace}
\newcommand{\qwen}{Qwen3-32B\xspace}
\newcommand{\deepseek}{DeepSeek-V3.1\xspace}
\newcommand{\claude}{Claude-4.6-Sonnet\xspace}
\newcommand{\llama}{Llama 4 Scout 17B\xspace}
\newcommand{\sota}{state-of-the-art\xspace}

% totalissues * 4     totalissues is below

% ============================================================
% Dataset Analysis: 
% ============================================================
% =========================
% Combined statistics by language
% =========================
% ---- Patch statistics ----
\newcommand{\PatchCountPy}{273}
\newcommand{\PatchCountCpp}{112}
\newcommand{\PatchCountJava}{109}

\newcommand{\PatchComplexityPy}{7.1}
\newcommand{\PatchComplexityCpp}{47.6}
\newcommand{\PatchComplexityJava}{19.2}

\newcommand{\PatchAddPy}{16.9}
\newcommand{\PatchAddCpp}{180.7}
\newcommand{\PatchAddJava}{74.8}

\newcommand{\PatchRemovePy}{7.7}
\newcommand{\PatchRemoveCpp}{82.6}
\newcommand{\PatchRemoveJava}{20.3}

% ---- Test statistics ----
\newcommand{\TestCountPy}{273}
\newcommand{\TestCountCpp}{112}
\newcommand{\TestCountJava}{109}

\newcommand{\TestComplexityPy}{7.2}
\newcommand{\TestComplexityCpp}{38.0}
\newcommand{\TestComplexityJava}{11.9}

\newcommand{\TestAddPy}{25.2}
\newcommand{\TestAddCpp}{277.8}
\newcommand{\TestAddJava}{72.2}

\newcommand{\TestRemovePy}{4.9}
\newcommand{\TestRemoveCpp}{17.5}
\newcommand{\TestRemoveJava}{2.0}

% ---- Bad Patch and Review statistics ----
\newcommand{\BadPatchCountPy}{164}
\newcommand{\BadPatchCountCpp}{44}
\newcommand{\BadPatchCountJava}{79}

\newcommand{\BadPatchComplexityPy}{2.870}
\newcommand{\BadPatchComplexityCpp}{3.641}
\newcommand{\BadPatchComplexityJava}{3.056}

\newcommand{\BadPatchAddPy}{3.909}
\newcommand{\BadPatchAddCpp}{5.455}
\newcommand{\BadPatchAddJava}{5.785}

\newcommand{\BadPatchRemovePy}{1.866}
\newcommand{\BadPatchRemoveCpp}{2.318}
\newcommand{\BadPatchRemoveJava}{1.861}

\newcommand{\ReviewSizePy}{253.6}
\newcommand{\ReviewSizeCpp}{319.6}
\newcommand{\ReviewSizeJava}{329.0}

% ============================================================
% Experiments: 
% ============================================================
% =========================
% SWE-Agent 
% =========================
% ---- Python ----
% Gemini 2.5 Flash
\newcommand{\PyGeminiBug}{38.1\%}
\newcommand{\PyGeminiTest}{14.0\%}
\newcommand{\PyGeminiReview}{29.9\%}
\newcommand{\PyGeminiStyle}{57.0\%}
\newcommand{\PyGeminiStyleOriginal}{72.2\%}
% DeepSeek V3.1
\newcommand{\PyDeepSeekBug}{56.4\%}
\newcommand{\PyDeepSeekTest}{18.7\%}
\newcommand{\PyDeepSeekReview}{52.2\%}
\newcommand{\PyDeepSeekStyle}{54.0\%}
\newcommand{\PyDeepSeekStyleOriginal}{73.4\%}
% GPT 5 mini
\newcommand{\PyGPTMiniBug}{47.3\%}
\newcommand{\PyGPTMiniTest}{6.2\%}
\newcommand{\PyGPTMiniReview}{30.5\%}
\newcommand{\PyGPTMiniStyle}{45.9\%}
\newcommand{\PyGPTMiniStyleOriginal}{56.3\%}
% Qwen3 32B
\newcommand{\PyQwenBug}{24.5\%}
\newcommand{\PyQwenTest}{4.0\%}
\newcommand{\PyQwenReview}{17.7\%}
\newcommand{\PyQwenStyle}{19.5\%}
\newcommand{\PyQwenStyleOriginal}{22.7\%}

% ---- C++ ----
% Gemini 2.5 Flash
\newcommand{\CppGeminiBug}{8.0\%}
\newcommand{\CppGeminiTest}{12.2\%}
\newcommand{\CppGeminiReview}{13.6\%}
\newcommand{\CppGeminiStyle}{30.5\%}
\newcommand{\CppGeminiStyleOriginal}{36.3\%}
% DeepSeek V3.1
\newcommand{\CppDeepSeekBug}{19.6\%}
\newcommand{\CppDeepSeekTest}{25.0\%}
\newcommand{\CppDeepSeekReview}{22.7\%}
\newcommand{\CppDeepSeekStyle}{18.8\%}
\newcommand{\CppDeepSeekStyleOriginal}{30.2\%}
% GPT 5 mini
\newcommand{\CppGPTMiniBug}{15.2\%}
\newcommand{\CppGPTMiniTest}{6.8\%}
\newcommand{\CppGPTMiniReview}{20.5\%}
\newcommand{\CppGPTMiniStyle}{19.5\%}
\newcommand{\CppGPTMiniStyleOriginal}{21.8\%}
% Qwen3 32B
\newcommand{\CppQwenBug}{3.8\%}
\newcommand{\CppQwenTest}{4.5\%}
\newcommand{\CppQwenReview}{4.5\%}
\newcommand{\CppQwenStyle}{6.7\%}
\newcommand{\CppQwenStyleOriginal}{8.6\%}

% ---- Java ----
% Gemini 2.5 Flash
\newcommand{\JavaGeminiBug}{14.7\%}
\newcommand{\JavaGeminiTest}{4.9\%}
\newcommand{\JavaGeminiReview}{31.6\%}
\newcommand{\JavaGeminiStyle}{22.0\%}
\newcommand{\JavaGeminiStyleOriginal}{60.4\%}
% DeepSeek V3.1
\newcommand{\JavaDeepSeekBug}{31.2\%}
\newcommand{\JavaDeepSeekTest}{20.9\%}
\newcommand{\JavaDeepSeekReview}{44.3\%}
\newcommand{\JavaDeepSeekStyle}{23.1\%}
\newcommand{\JavaDeepSeekStyleOriginal}{50.2\%}
% GPT 5 mini
\newcommand{\JavaGPTMiniBug}{22.0\%}
\newcommand{\JavaGPTMiniTest}{2.7\%}
\newcommand{\JavaGPTMiniReview}{26.6\%}
\newcommand{\JavaGPTMiniStyle}{22.1\%}
\newcommand{\JavaGPTMiniStyleOriginal}{25.0\%}
% Qwen3 32B
\newcommand{\JavaQwenBug}{10.1\%}
\newcommand{\JavaQwenTest}{1.3\%}
\newcommand{\JavaQwenReview}{15.2\%}
\newcommand{\JavaQwenStyle}{18.5\%}
\newcommand{\JavaQwenStyleOriginal}{23.3\%}

% =========================
% Claude 4.6 Sonnet + Gemini 2.5 Flash
% =========================
% Claude-4.6-Sonnet placeholders for SWE-Agent
\newcommand{\PyClaudeBug}{68.9\%}
\newcommand{\PyClaudeTest}{14.6\%}
\newcommand{\PyClaudeReview}{67.9\%}
\newcommand{\PyClaudeStyle}{61.2\%}

\newcommand{\CppClaudeBug}{8.0\%}
\newcommand{\CppClaudeTest}{13.6\%}
\newcommand{\CppClaudeReview}{9.1\%}
\newcommand{\CppClaudeStyle}{35.6\%}

\newcommand{\JavaClaudeBug}{15.6\%}
\newcommand{\JavaClaudeTest}{8.9\%}
\newcommand{\JavaClaudeReview}{20.3\%}
\newcommand{\JavaClaudeStyle}{27.3\%}

% =========================
% Aider + Gemini 2.5 Flash
% =========================
% ---- Python ----
\newcommand{\PyAiderBug}{32.4\%}
\newcommand{\PyAiderTest}{9.4\%}
\newcommand{\PyAiderReview}{26.8\%}
\newcommand{\PyAiderStyle}{48.6\%}
\newcommand{\PyAiderStyleOriginal}{60.3\%}
% ---- C++ ----
\newcommand{\CppAiderBug}{1.8\%}
\newcommand{\CppAiderTest}{2.3\%}
\newcommand{\CppAiderReview}{4.5\%}
\newcommand{\CppAiderStyle}{7.9\%}
\newcommand{\CppAiderStyleOriginal}{10.1\%}
% ---- Java ----
\newcommand{\JavaAiderBug}{19.3\%}
\newcommand{\JavaAiderTest}{3.9\%}
\newcommand{\JavaAiderReview}{25.3\%}
\newcommand{\JavaAiderStyle}{21.7\%}
\newcommand{\JavaAiderStyleOriginal}{60.9\%}

% =========================
% Style Review
% =========================
% Python — Fix Rate
\newcommand{\PyGeminiFix}{96.2\%}
\newcommand{\PyDeepSeekFix}{91.5\%}
\newcommand{\PyGPTMiniFix}{65.3\%}
\newcommand{\PyQwenFix}{30.5\%}
\newcommand{\PyAiderFix}{85.7\%}
% Python — Error Ratio
\newcommand{\PyGeminiErr}{0.377}
\newcommand{\PyDeepSeekErr}{0.299}
\newcommand{\PyGPTMiniErr}{0.457}
\newcommand{\PyQwenErr}{0.891}
\newcommand{\PyAiderErr}{0.482}
% Python — Raw Score
\newcommand{\PyGeminiRaw}{80.1\%}
\newcommand{\PyDeepSeekRaw}{79.5\%}
\newcommand{\PyGPTMiniRaw}{59.6\%}
\newcommand{\PyQwenRaw}{25.5\%}
\newcommand{\PyAiderRaw}{69.3\%}

% C++ — Fix Rate
\newcommand{\CppGeminiFix}{75.9\%}
\newcommand{\CppDeepSeekFix}{68.0\%}
\newcommand{\CppGPTMiniFix}{47.3\%}
\newcommand{\CppQwenFix}{35.3\%}
\newcommand{\CppAiderFix}{25.1\%}
% C++ — Error Ratio
\newcommand{\CppGeminiErr}{2.49}
\newcommand{\CppDeepSeekErr}{2.46}
\newcommand{\CppGPTMiniErr}{2.61}
\newcommand{\CppQwenErr}{2.87}
\newcommand{\CppAiderErr}{2.82}
% C++ — Raw Score
\newcommand{\CppGeminiRaw}{48.7\%}
\newcommand{\CppDeepSeekRaw}{41.4\%}
\newcommand{\CppGPTMiniRaw}{28.6\%}
\newcommand{\CppQwenRaw}{18.1\%}
\newcommand{\CppAiderRaw}{15.3\%}

% Java — Fix Rate
\newcommand{\JavaGeminiFix}{80.9\%}
\newcommand{\JavaDeepSeekFix}{77.9\%}
\newcommand{\JavaGPTMiniFix}{64.1\%}
\newcommand{\JavaQwenFix}{66.0\%}
\newcommand{\JavaAiderFix}{81.2\%}
% Java — Error Ratio
\newcommand{\JavaGeminiErr}{5.17}
\newcommand{\JavaDeepSeekErr}{5.46}
\newcommand{\JavaGPTMiniErr}{5.38}
\newcommand{\JavaQwenErr}{5.31}
\newcommand{\JavaAiderErr}{5.55}
% Java — Raw Score
\newcommand{\JavaGeminiRaw}{40.4\%}
\newcommand{\JavaDeepSeekRaw}{36.8\%}
\newcommand{\JavaGPTMiniRaw}{35.2\%}
\newcommand{\JavaQwenRaw}{34.5\%}
\newcommand{\JavaAiderRaw}{37.6\%}

% Claude-4.6-Sonnet placeholders for style analysis
\newcommand{\PyClaudeFix}{71.6\%}
\newcommand{\PyClaudeErr}{0.368}

\newcommand{\CppClaudeFix}{73.2\%}
\newcommand{\CppClaudeErr}{2.13}

\newcommand{\JavaClaudeFix}{78.5\%}
\newcommand{\JavaClaudeErr}{5.22}

% =========================
% Patch Generation Rate
% =========================
% ---- Python ----
\newcommand{\PyGeminiPatchBug}{93.8\%}
\newcommand{\PyGeminiPatchTest}{92.0\%}
\newcommand{\PyGeminiPatchReview}{92.7\%}
\newcommand{\PyGeminiPatchStyle}{91.7\%}

\newcommand{\PyDeepSeekPatchBug}{96.3\%}
\newcommand{\PyDeepSeekPatchTest}{94.8\%}
\newcommand{\PyDeepSeekPatchReview}{95.1\%}
\newcommand{\PyDeepSeekPatchStyle}{93.8\%}

\newcommand{\PyGPTMiniPatchBug}{76.2\%}
\newcommand{\PyGPTMiniPatchTest}{64.8\%}
\newcommand{\PyGPTMiniPatchReview}{61.0\%}
\newcommand{\PyGPTMiniPatchStyle}{69.3\%}

\newcommand{\PyQwenPatchBug}{79.1\%}
\newcommand{\PyQwenPatchTest}{90.8\%}
\newcommand{\PyQwenPatchReview}{78.0\%}
\newcommand{\PyQwenPatchStyle}{35.7\%}

% ---- C++ ----
\newcommand{\CppGeminiPatchBug}{98.2\%}
\newcommand{\CppGeminiPatchTest}{97.7\%}
\newcommand{\CppGeminiPatchReview}{77.3\%}
\newcommand{\CppGeminiPatchStyle}{80.3\%}

\newcommand{\CppDeepSeekPatchBug}{96.4\%}
\newcommand{\CppDeepSeekPatchTest}{75.0\%}
\newcommand{\CppDeepSeekPatchReview}{97.7\%}
\newcommand{\CppDeepSeekPatchStyle}{85.7\%}

\newcommand{\CppGPTMiniPatchBug}{62.5\%}
\newcommand{\CppGPTMiniPatchTest}{54.5\%}
\newcommand{\CppGPTMiniPatchReview}{56.8\%}
\newcommand{\CppGPTMiniPatchStyle}{48.3\%}

\newcommand{\CppQwenPatchBug}{70.5\%}
\newcommand{\CppQwenPatchTest}{97.7\%}
\newcommand{\CppQwenPatchReview}{88.6\%}
\newcommand{\CppQwenPatchStyle}{47.6\%}

% ---- Java ----
\newcommand{\JavaGeminiPatchBug}{99.1\%}
\newcommand{\JavaGeminiPatchTest}{79.2\%}
\newcommand{\JavaGeminiPatchReview}{93.7\%}
\newcommand{\JavaGeminiPatchStyle}{84.7\%}

\newcommand{\JavaDeepSeekPatchBug}{93.6\%}
\newcommand{\JavaDeepSeekPatchTest}{87.0\%}
\newcommand{\JavaDeepSeekPatchReview}{91.1\%}
\newcommand{\JavaDeepSeekPatchStyle}{91.1\%}

\newcommand{\JavaGPTMiniPatchBug}{45.9\%}
\newcommand{\JavaGPTMiniPatchTest}{45.5\%}
\newcommand{\JavaGPTMiniPatchReview}{51.9\%}
\newcommand{\JavaGPTMiniPatchStyle}{50.0\%}

\newcommand{\JavaQwenPatchBug}{75.2\%}
\newcommand{\JavaQwenPatchTest}{90.9\%}
\newcommand{\JavaQwenPatchReview}{77.2\%}
\newcommand{\JavaQwenPatchStyle}{44.4\%}

% ---- Claude ----
% Claude-4.6-Sonnet patch generation rates
\newcommand{\PyClaudePatchBug}{94.1\%}
\newcommand{\PyClaudePatchTest}{60.4\%}
\newcommand{\PyClaudePatchReview}{81.7\%}
\newcommand{\PyClaudePatchStyle}{70.7\%}

\newcommand{\JavaClaudePatchBug}{27.5\%}
\newcommand{\JavaClaudePatchTest}{20.3\%}
\newcommand{\JavaClaudePatchReview}{30.4\%}
\newcommand{\JavaClaudePatchStyle}{75.0\%}

\newcommand{\CppClaudePatchBug}{61.6\%}
\newcommand{\CppClaudePatchTest}{27.3\%}
\newcommand{\CppClaudePatchReview}{81.8\%}
\newcommand{\CppClaudePatchStyle}{35.6\%}

\newcommand{\Fix}[1]{\textcolor{red}{#1}}
\newcommand{\totalprojects}{19\xspace}
\newcommand{\totalissues}{728\xspace}

\newcommand{\totaltasks}{1794\xspace}

\newcommand{\pythonbftasks}{273\xspace}
\newcommand{\javabftasks}{109\xspace}
\newcommand{\cppbftasks}{112\xspace}

\newcommand{\totalbftasks}{494\xspace}

\newcommand{\pythontasks}{963\xspace}
\newcommand{\javatasks}{418\xspace}
\newcommand{\cpptasks}{413\xspace}

\newcommand{\pythonsftasks}{144\xspace}
\newcommand{\javasftasks}{124\xspace}
\newcommand{\cppsftasks}{147\xspace}

\newcommand{\totalrepos}{28\xspace}